\documentclass[10pt,twocolumn,letterpaper]{article}

\usepackage{iccv}              

\usepackage{booktabs}
\usepackage{multicol}
\usepackage{colortbl}
\usepackage{multirow}
\usepackage{amsthm}
\newtheorem{theorem}{Theorem} 
\newtheorem{lemma}{Lemma} 
\newtheorem{remark}{Remark}
\usepackage{amssymb}
\usepackage{pifont}
\usepackage{comment}
\usepackage[normalem]{ulem}
\usepackage{xcolor}
\usepackage{amsthm,amsmath,amssymb,,bm,bbm}
\usepackage{mdframed}
\usepackage{multicol} 
\theoremstyle{plain}
\newtheorem{prop}[theorem]{Proposition}
\theoremstyle{definition}
\usepackage{makecell}

\newmdtheoremenv{definition}[theorem]{Definition}

\theoremstyle{remark}
\newmdtheoremenv{corollary}{Corollary}[theorem]
\definecolor{iccvblue}{rgb}{0.21,0.49,0.74}
\usepackage[pagebackref,breaklinks,colorlinks,allcolors=iccvblue]{hyperref}

\def\paperID{11491} 
\def\confName{ICCV}
\def\confYear{2025}

\title{Cracking Instance Jigsaw Puzzles: An Alternative to Multiple Instance Learning for Whole Slide Image Analysis}

\author{ Xiwen Chen$^{1}$\thanks{Equal contribution} \enspace Peijie Qiu$^{2}$\footnotemark[1] \enspace Wenhui Zhu$^{3}$\footnotemark[1] \enspace Hao Wang$^{1}$ \enspace Huayu Li $^{4}$\enspace   Xuanzhao Dong$^{3}$\enspace \\ Xiaotong Sun$^{5}$  \enspace  Xiaobing Yu $^{2}$ Yalin Wang$^{3}$ \enspace Abolfazl Razi$^{1}$ \enspace Aristeidis Sotiras$^{2}$\thanks{\textit{Corresponding Author}}\\
$^{1}$ Clemson University, 
$^{2}$ Washington University in St. Louis, 
$^{3}$ Arizona State University, \\
$^{4}$ University of Arizona, 
$^{5}$ University of Arkansas
}

\begin{document}
\maketitle

\begin{abstract}
    While multiple instance learning (MIL) has shown to be a promising approach for histopathological whole slide image (WSI) analysis, its reliance on permutation invariance significantly limits its capacity to effectively uncover semantic correlations between instances within WSIs. 
    Based on our empirical and theoretical investigations, we argue that approaches that are not permutation-invariant but better capture spatial correlations between instances can offer more effective solutions. In light of these findings, we propose a novel alternative to existing MIL for WSI analysis by learning to restore the order of instances from their randomly shuffled arrangement. We term this task as cracking an instance jigsaw puzzle problem, where semantic correlations between instances are uncovered. To tackle the instance jigsaw puzzles, we propose a novel Siamese network solution, which is theoretically justified by optimal transport theory. We validate the proposed method on WSI classification and survival prediction tasks, where the proposed method outperforms the recent state-of-the-art MIL competitors. The code is available at \url{https://github.com/xiwenc1/MIL-JigsawPuzzles}.
\end{abstract}

\section{Introduction}
\begin{figure}
    \centering
    \includegraphics[width=0.9\linewidth]{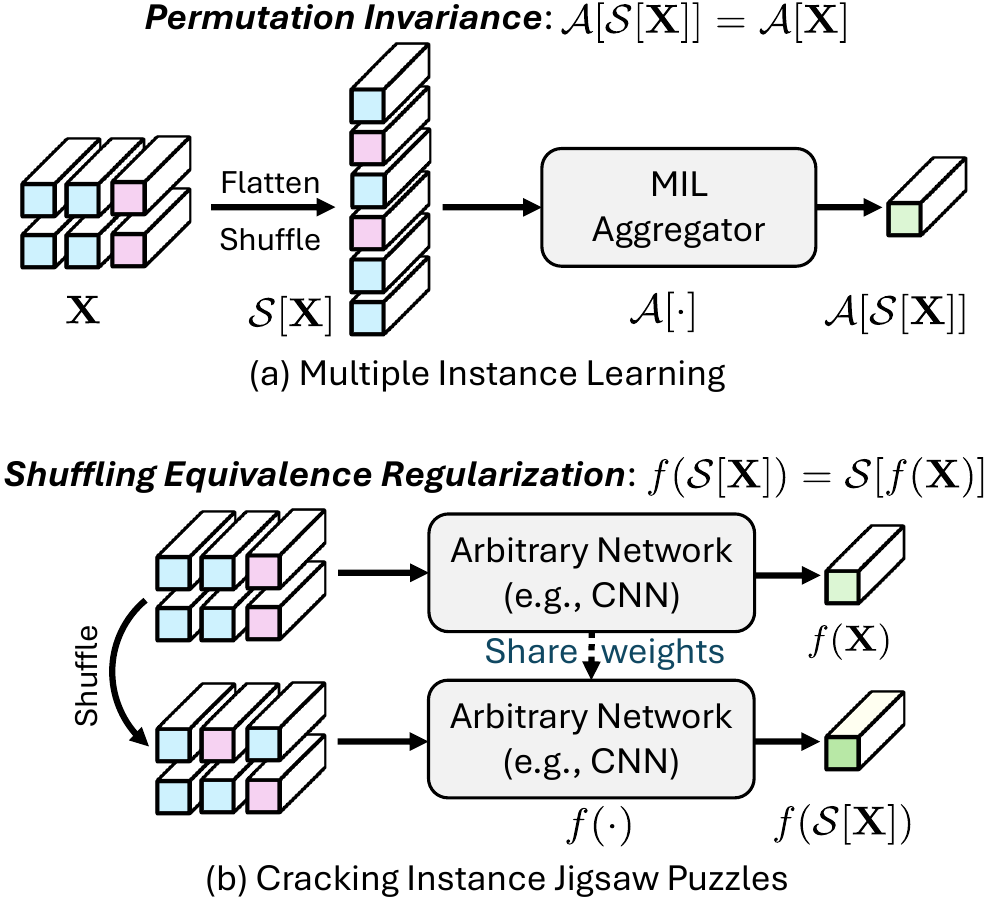}
    \caption{An illustrative comparison between (a) the traditional MIL method and (b) the proposed method for solving instance jigsaw puzzles. Compared to MIL, the proposed method is advantageous in uncovering semantic correlations between instances.}
    \label{fig:graphical_abstract}
\end{figure}

Histopathological whole slide images (WSIs) are essential in modern digital pathology, supporting tasks such as cancer diagnosis~\cite{cancer,cancer2} and survival prediction~\cite{jaume2024modeling,song2024multimodal}. However, the gigapixel resolution of WSIs poses significant computational challenges, hindering the direct application of traditional deep learning methods (\textit{e.g.,} convolution neural networks) to WSI analysis. To mitigate this issue, the prevailing studies~\cite{li2021dual,shao2021transmil,zhang2022dtfd,clam-sb,lowrankmil,zhu2025how} for WSI analysis rely on the multiple instance learning (MIL). In this process, a WSI (\textit{a.k.a.,} bag in the context of MIL) is first patchified into smaller (\textit{e.g.,} $224 \times 224$) tiles/instances, and features are then extracted using pre-trained feature extractors (\textit{e.g.,} ResNet~\cite{resnet}). MIL acts as an aggregator, combining features from individual tiles, which are then fed into a simple MLP (\textit{e.g.,} a linear classifier in most cases) to output the final prediction. However, MIL aggregator designs are largely constrained by the requirement of permutation invariance (see Fig.~\ref{fig:graphical_abstract}(a)), a fundamental assumption of the orthodox MIL problem\footnote{Technically, methods that violate the permutation-invariance assumption are not considered MIL~\cite{ilse2018attention,raff2023reproducibility}.}~\cite{ilse2018attention,raff2023reproducibility}. \textit{In what follows, we will investigate the necessity of this permutation invariance (or MIL) from both empirical and theoretical perspectives.}

Our primary motivation is that, although permutation invariance is a fundamental assumption for MIL problems, it overlooks the spatial correlations between tiles in WSIs—an essential ingredient for the success of most image-driven tasks. In addition, there is a large consensus that neighboring tiles in WSIs are likely to belong to the same tissue category~\cite{song2024morphological}. However, a dilemma immediately arises between maintaining permutation invariance and preserving spatial correlations. Let us take the convolutional operation as an example due to its popularity and success in capturing spatial correlations. A convolutional operation is inherently \textit{not} permutation-invariant since the alternation of spatial arrangement leads to entirely different outputs. However, we observe that injecting some convolutional layers into many MIL aggregators leads to immediate performance gain. More interestingly, directly replacing commonly used MIL aggregators with a simple CNN even achieves a better performance, which has surprisingly not been explored in the literature. We kindly direct the readers to Sec.~\ref{sec:3.1} for detailed analyses. These facts lead us to question \textit{whether we really need an MIL for WSI analysis}. Instead, we argue that, \textit{despite violating the permutation-invariant constraint, modeling the spatial correlations between tiles/instances is necessary}.

Another subtle but important fact is that most MIL methods shuffle the instances before feeding them into the MIL aggregators to consolidate the permutation-invariant requirement further (see Fig.~\ref{fig:graphical_abstract}(a)). This explicitly destroys the spatial correlations between tiles, which makes it even harder for standard MILs to capture spatial correlations. Although MIL methods that leverage non-local information (\textit{e.g.,} attention mechanisms~\cite{li2021dual,shao2021transmil,lowrankmil}) partially mitigate this problem, they mainly capture the similarities between latent features across tiles but do not leverage spatial information of tiles. Due to the fact that the feature space is quite noisy due to the suboptimal two-stage learning scheme, the learned correlations between tiles in the feature space are spurious~\cite{shao2021transmil,lin2023interventional}. Graph Neural Network (GNN)-based MILs~\cite{chen2021whole,zheng2022graph,hou2022h,chan2023histopathology} offer an alternative approach to capturing spatial relationships among instances with violated the permutation-invariance assumption. However, they often require additional knowledge (\textit{e.g.,} patch connectivity), which may not always be accessible, to construct the graph.


Rather than achieving better modeling instance correlation through different attention mechanisms, naive position encoding (\textit{e.g.,} \texttt{sin-cos}), or accessing additional information, we discover that learning to restore the original order from shuffled tiles can lead to better performance by exploiting their semantic correlations. We term this process as solving an instance jigsaw puzzle
\footnote{Our instance jigsaw puzzles problem fundamentally differs from the one in~\cite{noroozi2016unsupervised}. The core difference is that we formulate instance jigsaw puzzles as a regularization instead of an image representation learning method. As a result, the solution to the instance jigsaw puzzles is different from the one in~\cite{noroozi2016unsupervised} (see~\Cref{sec:4.2,sec:4.3} for a detailed discussion).} 
by drawing a connection to the well-known jigsaw puzzle problem in the context of self-supervised representation learning~\cite{noroozi2016unsupervised}, which is proven to capture semantically meaningful representations. To solve the instance jigsaw puzzle problem, we propose a novel \textit{shuffling equivalence regularization} loss implemented as a Siamese network  (see Fig.~\ref{fig:graphical_abstract}(b)). We further justify the proposed solution from the lens of optimal transport theory, showing its equivalence to minimizing the optimal transport cost required to restore the original order of tiles from their shuffled arrangement. We empirically validate the effectiveness of the proposed method on WSI classification and survival prediction tasks.

In summary, our contributions are three-fold: \textbf{(i)} We propose a superior alternative to existing MIL for whole slide image analysis by solving an instance jigsaw puzzle, which can leverage the spatial correlations between tiles without a permutation-invariant constraint. The relaxation of permutation invariance enables more flexible choices of networks that can better handle spatial correlations (\textit{e.g.,} CNNs).  \textbf{(ii)} We propose a novel solution to the instance jigsaw puzzle problem by leveraging the Siamese network and optimal transport theory. \textbf{(iii)} Empirical results on WSI classification and survival prediction tasks show the superiority of the proposed method to traditional MIL methods.


\section{Related Works}\label{sec:2}
To the best of our knowledge, the proposed method is fundamentally different from any existing method for WSI analysis while maintaining relevance to MIL-based methods. Hence, we provide a brief review of MIL applications to WSI analysis and their connections to our method. Previous MIL methods revolve around enhancing either the MIL aggregators by capturing instance correlations~\cite{shao2021transmil,zhang2022dtfd,li2021dual,lowrankmil,tang2023multiple,zhu2025dgr,zhang2023attention} or the instance-level feature extractors through self-supervised learning~\cite{li2021dual,lowrankmil,wang2021transpath}. Since our method is more related to MIL aggregators, we focus on this direction hereafter. 
In contrast, self-supervised learning offers a complementary approach to MIL aggregators and our method, enhancing patch-level WSI embeddings for both.

The previous MIL aggregators for WSI analysis can be roughly divided into instance-level and bag-level MIL aggregators. However, the instance-level MIL aggregators~\cite{zhu2025dgr,hou2016patch,coudray2018classification,wang2019weakly,chikontwe2020multiple,lin2022interventional} empirically show inferior performance compared to their bag-level counterparts, because propagating bag-level labels to their instances inevitably introduces noisy instance-level supervision~\cite{wang2018revisiting,li2021dual}. Therefore, we focus on bag-level MIL methods~\cite{wang2018revisiting,campanella2019clinical,ilse2018attention,shao2021transmil,zhang2022dtfd,li2021dual,lowrankmil,tang2023multiple,zhu2025dgr,zhang2023attention} hereafter. Since the introduction of MILNet~\cite{wang2018revisiting}, deep neural networks have dominated the design of bag-level MIL methods. ABMIL~\cite{ilse2018attention} further expands its interpretability by leveraging an attention pooling mechanism, which has now become the de-facto standard for WSI analysis. However, ABMIL treats each instance independently without considering their inherent correlations. Therefore, many of its follow-up works focus on addressing this issue by leveraging simple instance clustering~\cite{clam-sb}, recurrent neural network~\cite{campanella2019clinical}, self-attention~\cite{shao2021transmil,lowrankmil,xiong2023diagnose}, cross-attention~\cite{zhu2025dgr}, prototypical dictionary representations~\cite{qiu2023sc,lin2023interventional}, and graph representations~\cite{chen2021whole,zheng2022graph,hou2022h,bontempo2023graph}. There is another line of work that enhances the MIL methods by leveraging data augmentation, such as pseudo bag~\cite{zhang2022dtfd}, mixup~\cite{yang2022remix,chen2023rankmix}, and stochastic instance masking~\cite{qu2023boosting,tang2023multiple,zhang2023attention}.

Similar to the above MIL methods, the proposed method operates on instances instead of raw WSI slides due to computational intractability. \textit{However, our method is not classified as an MIL since it does not satisfy the permutation-invariant requirement.} Instead, our method enables a more flexible choice of neural network structures that can explicitly capture spatial correlations (i.e., mostly position prior) between instances (\textit{e.g.,} CNNs, which has yet to be explored previously). Although previous MIL methods that model instance correlations partially capture the spatial information by comparing the similarities between instances in the feature space (\textit{e.g.,} through Transformers~\cite{shao2021transmil} and graphs~\cite{hou2022h,chan2023histopathology}), they are either prone to capture spurious correlations due to the noisy feature space and cannot be generalized to any network architecture, or relies on addition knowledge to construct the graph. Instead, our method leverages the instance ordering information to uncover their semantics without the dependence on extra information, \textit{e.g.,} patch connectivity~\cite{zheng2022graph,chen2021whole}.



\section{Rethink MIL for WSI Analysis}
In this section, we provide some new insights into applying MIL for WSI classification through theoretical analyses and empirical observations.

\subsection{WSI analysis as an MIL problem}

\noindent\textbf{MIL Formulation.}
Without loss of generality, a WSI slide (\textit{a.k.a.,} a bag) can be represented as $\bm{\mathrm{X}} = \{\bm{\mathrm{x}}_1, \bm{\mathrm{x}}_2, \cdots, \bm{\mathrm{x}}_n\}$, where $\boldsymbol{x}_i$ denotes the $i$-th tile/ instance. Our goal is to learn a scoring function $S: \mathcal{X} \rightarrow \mathcal{Y}$ that maps the instance space $\mathcal{X}$ to bag-level label space $\mathcal{Y}$. The standard MIL problem defines that a bag is classified as positive if and only if it contains at least one positive instance: 
\begin{equation}\nonumber
    \bm{\mathrm{Y}} = \begin{cases}
     0, \ \text{iff} \ \sum_{i} y_i = 0 \\
     1, \ \text{otherwise},
    \end{cases}
\end{equation}
where $y_i \in \{0, 1\}$ is the instance-level label for the $i$-th instance, and $\bm{\mathrm{Y}} \in \{0, 1\}$ is the bag-level label. The instance-level label $y_i$ is typically unknown to us in the regime of WSI classification. In practice, we typically consider each instance $\bm{\mathrm{x}}_i$
as a $d$-dimensional feature vector extracted by some pre-trained backbone networks. Although this procedure is suboptimal, back-propagating through thousands of input image patches during a single mini-batch optimization is computationally expensive when operating on raw WSIs.

\noindent\textbf{MIL Solution.} In practice, a MIL model outputs a probabilistic bag-level prediction $\hat{Y} \in [0, 1]$ through a scoring function $ P(\bm{\mathrm{X}})$. For the MIL problem to be valid, the scoring function $P(\cdot)$ needs to be permutation-invariant, which is achieved if and only if it can be approximated in the following form (see~\citep[][Theorem 2]{ilse2018attention} and ~\citep[][Theorem 1]{shao2021transmil}):
\begin{equation}\nonumber
    |P(\bm{\mathrm{X}}) - g(\underset{\bm{\mathrm{x}}\in \bm{\mathrm{X}}}{\mathcal{A}} [\bm{\mathrm{x}}]) | < \epsilon, \quad \forall \epsilon > 0,
\end{equation}
where $g(\cdot)$ can be any continuous functions, and the operator $\mathcal{A}[\cdot]$ is an permutation-invariant aggregation (a.k.a. MIL pooling) operator. 
In what follows, we restrict ourselves to the MIL problem parameterized by deep neural networks (which we define as DeepMIL for brevity) due to its popularity and success in WSI classification. In this case, $\mathcal{A}_\theta(\cdot)$ is parameterized by a network, and $g_\phi(\cdot)$ is a linear classifier with parameters $\phi$. 
It is worth noting that $\mathcal{A}_\theta(\cdot)$ may also include some embedding layers at the very beginning to reduce the dimensionality of initial feature embeddings. 


\noindent\textbf{Dive into Permutation Invariance.} Permutation invariance is a unique feature of MIL, which restricts the network designs of most previous DeepMIL models~\cite{wang2018revisiting,ilse2018attention,li2021dual,shao2021transmil}.
Although non-parametric MIL aggregators (\textit{e.g.,} \texttt{max}-operator, \texttt{mean}-operator, \texttt{log-sum-exp}-operator) are considered in the early literature~\citep[][]{wang2018revisiting}, more modern MIL methods leverage parametric attention-based aggregators~\cite{ilse2018attention,shao2021transmil} to gain better performance and interpretability.
To the best of our knowledge, their permutation-invariant property has not been formally discussed previously. Here, we provide a formal discussion in Prop.~\ref{prop:1}.
\begin{prop}\label{prop:1}
    Attention pooling in ABMIL and TransMIL (w/o positional encoding) is permutation-invariant. However, introducing positional encoding into permutation-invariant MIL aggregators destroys this invariance.
\end{prop}
\begin{proof}
    We defer the formal proof to \textcolor{Periwinkle} {\bf Appendix A.1}.
\end{proof}

\subsection{Importance of Spatial Information in MIL}\label{sec:3.1}

\begin{figure}[!t]
    \centering
    \includegraphics[width=0.98\linewidth]{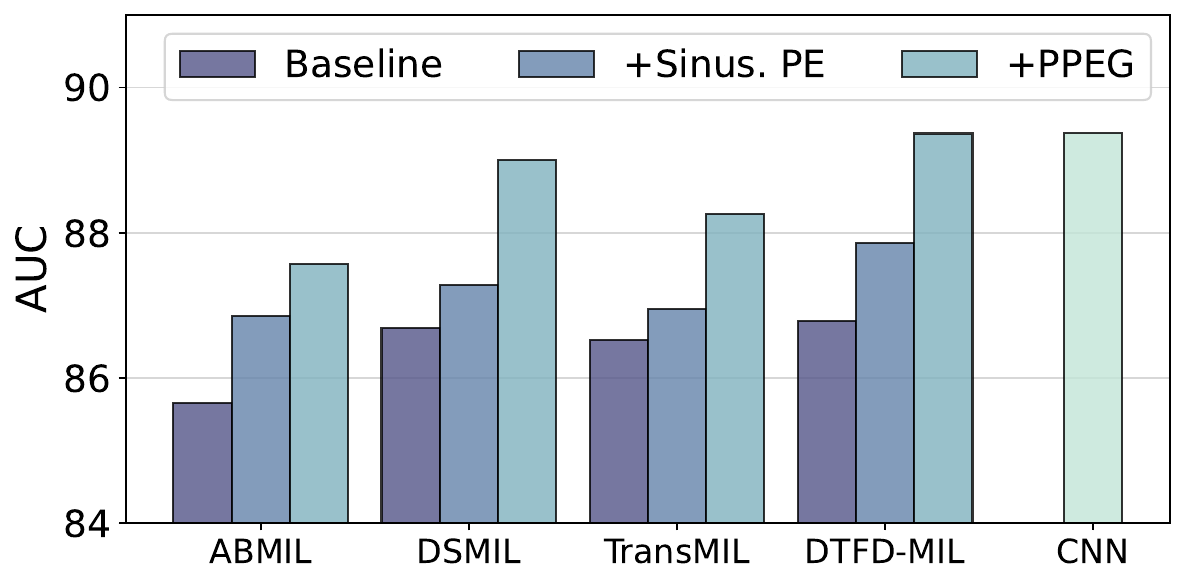}
    \vspace{-0.1in}
    \caption{A comparison of performance between four different MIL aggregators (w/ and w/o positional encoding) and a CNN on CAMELYON16 dataset using features extracted by a Swin-Transformer. \texttt{Sinus.~PE} denotes sinusoidal positional encoding, and PPEG denotes pyramid positional encoding generator.}
    \label{fig:inv1}
    \vspace{-0.1in}
\end{figure}

\noindent \textbf{Empirical Observation.} We conduct comparative studies using four popular MIL aggregators (\textit{i.e.,} ABMIL~\cite{ilse2018attention}, DSMIL~\cite{li2021dual}, DTFD-MIL~\cite{zhang2022dtfd}, TransMIL~\cite{shao2021transmil}) as well as a simple CNN on CAMELYON16 dataset.
Based on the results shown in~\Cref{fig:inv1}, we have the following two key observations: \textbf{(i)} Introducing positional encoding (\textit{e.g.,} sinusoidal positional encoding~\cite{vaswani2017attention} and pyramid positional encoding generator~\citep[PPEG,][]{shao2021transmil}) into MIL aggregators consistently leads to performance gain. 
\textbf{(ii)} A simple CNN can perform as well as the best MIL aggregator but outperforms all plain MIL aggregators w/o positional encoding. 

Concretely, despite violating most MIL assumptions~\citep{raff2024reproducibility}) including permutation-invariant requirement (see Prop.~\ref{prop:1}), DTFD-MIL\footnote{A variant of ABMIL with pseudo-bag data augmentation and knowledge distillation.} with PPEG (technically not an orthodox MIL method) performs the best. Even when we intentionally destroy the permutation invariance in ABMIL with positional encoding, we still observe a performance improvement. 
These findings coincide with observations in MIL for time-series classification~\cite{early2024inherently,chen2024timemil}, where the introduction of positional encoding into MIL also leads to performance gain. 

\textit{These facts question the necessity of orthodox MIL, which strictly follows the standard MIL assumptions, for WSI analysis. Could explicitly learning semantics among instances offer a more effective solution ?}

\noindent \textbf{Justification.} We conjecture the effectiveness of positional information in boosting the performance of prior MIL aggregators stems from the fact that neighboring instances are likely to belong to the same category. Although this biological phenomenon has been consistently observed across various pathological subtyping studies~\cite{shao2021transmil,zhu2025dgr,clinicdiversity,clinicdiversity2,clinicdiversity3,clinicdiversity4,clinicdiversity5}, it is not fully explored in the context of WSI analysis. The PPEG, implemented as a pyramid of convolutional layers (\textit{i.e.,} a special type of conditional positional encoding~\cite{chu2021conditional}), partially utilizes this spatial prior by leveraging the inductive biases of convolution, such as locality and shift-invariant. Furthermore, a full convolution network with stronger spatial inductive biases can perform similarly.

Below, we provide a theoretical justification of the effectiveness of positional information for MIL. The intuition is that incorporating additional information does not negatively impact the model's performance.





\begin{theorem}
Incorporating positional information can lower the classification-error upper bound. This is because
\begin{align}
   H(\boldsymbol{\mathrm{Y}}|\boldsymbol{\mathrm{X}}) \geq H(\boldsymbol{\mathrm{Y}}|\boldsymbol{\mathrm{X}}, \boldsymbol{\mathrm{P}}), 
\end{align}
where $\boldsymbol{\mathrm{P}}= \{\boldsymbol{\mathrm{p}}_1, \boldsymbol{\mathrm{p}}_2, \cdots, \boldsymbol{\mathrm{p}}_n\}$ denotes the positional coordinates associated with each instance on the raw WSIs.
\end{theorem}
\begin{proof}
    Due to the monotonicity of conditional information, conditioning on additional positional information only enables preserving or reducing the uncertainty on $\boldsymbol{Y}$. We postpone the complete proof to \textcolor{Periwinkle} {\bf Appendix A.2}.
\end{proof}

\section{Cracking Instance Jigsaw Puzzles}
In this section, we introduce a novel method, termed solving instance jigsaw puzzles, which enforces the network to learn the semantics among instances.


\noindent \textbf{Instance Jigsaw Puzzles.} Although the PPEG used in TransMIL~\cite{shao2021transmil} also imposes positional information into the MIL framework, it mainly relies on convolutional operation, which is typically limited by its respective field. As a consequence, it mainly discoverers the local semantics/similarities among instances while needing a globally contextual understanding of the semantics among instances. The convolutional operation is particularly ineffective if the input instances are shuffled in most MIL practices, where the local structures of instances are not preserved anymore. 

In contrast, unsupervised representation learning methods are advantageous in discovering semantic representations by establishing a pretext task. One particular pretext task, termed solving jigsaw puzzles~\cite{noroozi2016unsupervised}, involves learning semantic representations by randomly shuffling patchified tiles of an image and then predicting their correct positions through a neural network. We realize that this method has a tied connection to MIL for WSI analysis, which surprisingly has not been explored in previous literature. Specifically, the input instances to MIL aggregators are typically shuffled to pursue permutation invariance. The only difference is that the previous MIL methods for WSI analysis directly use the shuffled instances for classification (see Fig.~\ref{fig:graphical_abstract}(a)) rather than predicting its correct locations. We define the task of shuffling instances and then restoring their original positions as solving \textit{instance jigsaw puzzles} problem. We argue that solving this instance of a jigsaw puzzle can help MIL learn semantics among instances effectively. 

\subsection{A Siamese Network Solution}\label{sec:4.2}
To solve the above instance jigsaw puzzles problem, we propose a Siamese network solution. Specifically, given a shuffling operator $\mathcal{S}[\cdot]$, which randomly shuffles input instances, the following equivalence should hold for solving the instance jigsaw puzzles:
\begin{equation}
    f_\theta(\mathcal{S}[\bm{\mathrm{X}}]) = \mathcal{S}[f_\theta(\bm{\mathrm{X}})],
\end{equation}
where $f_\theta$ is parameterized by a neural network, the rationale is that if a network can learn to restore the correct arrangement from shuffled instances, applying the inverse shuffling operation $\mathcal{S}^{-1}[\cdot]$ to the network's output should recover the original arrangement: $\mathcal{S}^{-1}[f_\theta(\mathcal{S}[\bm{\mathrm{X}}])] = f_\theta(\bm{\mathrm{X}})$. To improve readability, we postpone the theoretical justification to Sec.~\ref{sec:4.3}.
Accordingly, we design a shuffling equivalence regularization loss as follows:
\begin{equation}\label{eq:equv_loss}
    \mathcal{L}_{\text{Equv}}(\bm{\mathrm{X}}) = \frac{1}{2n} || f(\mathcal{S}[\bm{\mathrm{X}}]) - \mathcal{S}[f(\bm{\mathrm{X}})] ||_2^2, 
\end{equation}
which penalizing the mean squared  error between $f(\mathcal{S}[\bm{\mathrm{X}}])$ and $\mathcal{S}[f(\bm{\mathrm{X}})]$. This equivalence loss is then implemented as a Siamese network with two branches that share the same weights (see Fig.~\ref{fig:graphical_abstract}(b)). The first branch takes as input the unshuffled instances $\bm{\mathrm{X}}$, while the second branch takes as input the shuffled instances $\mathcal{S}[\bm{\mathrm{X}}]$. 
The final objective for WSI classification is then a weighted combination of the equivalence loss and the binary cross-entropy loss ($\mathcal{L}_{\text{BCE}}$):
\begin{equation}\label{eq:weight}
    \mathcal{L}_{\text{final}}(\bm{\mathrm{X}}, \bm{\mathrm{Y}}) = \mathcal{L}_{\text{BCE}}(\bm{\mathrm{X}}, \bm{\mathrm{Y}}) + \lambda \mathcal{L}_{\text{Equv}}(\bm{\mathrm{X}}),
\end{equation}
Where $\lambda$ controls the regularization strength, we replace the cross entropy loss with discrete-time negative log-likelihood loss used in~\cite{chen2021whole} for the survival prediction task.

\noindent \textbf{Network Design.}
The relaxation of the permutation-invariant constraint enables us to employ any neural network in our proposed method. As a proof of concept, we validate our method using both transformers and CNN due to their success in vision tasks. 
Specifically, we employ the same transformer architecture outlined in TransMIL~\cite{shao2021transmil} for a fair comparison. 
We would also like to point out that a plain transformer without positional encoding may result in a trivial solution. Since it is permutation-invariant, learning the semantic information from restoring the shuffled arrangement is impossible. The CNN is implemented as a stack of 2D convolutional residual blocks. Unlike transformers, which process 1D inputs, 2D CNNs operate on 2D inputs. Therefore, we reshape the original 1D instance embeddings into their 2D form with the necessary padding. After feature extraction, we apply a \texttt{global average pooling} followed by a linear classifier to both CNN and transformer to obtain bag-level predictions. We opt for this choice as it can be simply implemented in both architectures
and allows us to easily leverage the \textit{class activation map (CAM)}~\cite{zhou2016learning} for interpretability (see \textcolor{Periwinkle} {\bf Appendix B.2}).
As the architectural design is not the focus of this paper, detailed network architectures are deferred to \textcolor{Periwinkle} {\bf Appendix B.1}. Although the most naive implementation of the proposed Siamese network almost doubles the computation compared to the single-branch MIL frameworks, we mitigate this issue by leveraging parallelization (see \textcolor{Periwinkle} {\bf Appendix B.3}).


\subsection{An Optimal Transport Theory Justification}\label{sec:4.3}
The proposed equivalence regularization loss is justified by Optimal Transport (OT) theory. Restoring the correct ordering of a set of positions  $\boldsymbol{\mathrm{P}}= \{\boldsymbol{\mathrm{p}}_1, \boldsymbol{\mathrm{p}}_2, \cdots, \boldsymbol{\mathrm{p}}_n\}$ from its randomly shuffled arrangement $\boldsymbol{\mathrm{P}}' = \mathcal{S}[\boldsymbol{\mathrm{P}}]$ can be formulated as finding an optimal transport plan $\bm{\mathrm{T}}_\#$ by minimizing a known cost function of moving all the elements form $\boldsymbol{\mathrm{P}}'$ to $\boldsymbol{\mathrm{P}}$, or vice versa. This can be achieved by minimizing the earth mover's distance with a linear cost (see \textcolor{Periwinkle} {\bf Appendix C.1}).
However, solving this forward OT problem requires an iterative Sinkhorn algorithm~\cite{cuturi2013sinkhorn,peyre2019computational}, which is not computationally friendly to neural networks. 

A key observation is that we have access to the transport plan, \textit{i.e.,} the inverse shuffling operation $\mathcal{S}^{-1}$. Although this may not be the optimal transport plan, it still serves our purpose of restoring the ordering of the original instances with negligible computational cost. However, the cost function in our case is more complex and unknown, as it must account for the instance embeddings $\bm{\mathrm{X}}$ to capture meaningful semantics between instances. 
This shifts our main goal to find an optimal cost measure between $\bm{\mathrm{x}}_i$ and $\bm{\mathrm{x}}'_i$ under a known transport plan.
This problem is formally defined as an \textit{inverse optimal transport} problem~\cite{stuart2020inverse,li2019learning}:
\begin{equation}
    \min\limits_{\theta} \mathcal{L}(\Tilde{\bm{\mathrm{T}}}[c_\theta], \bm{\mathrm{T}}_\#[c_\theta]), 
\end{equation}
where $c_\theta$ is the parameterized cost function, and $\mathcal{L}(\cdot)$ is a loss function, \textit{e.g.,} MSE. $\Tilde{\bm{\mathrm{T}}}$ and $\bm{\mathrm{T}}_\#$ are the observed and optimal transport plan. However, solving this inverse OT problem requires solving the forward OT problem to obtain $\bm{\mathrm{T}}_\#$~\cite{stuart2020inverse,li2019learning}, which is computationally expensive. Therefore, we approximate the optimal transport plan with the observed one to solve the inverse OT problem efficiently.

\begin{theorem}
    When approximating the optimal transport plan $\bm{\mathrm{T}}_\#$ with the inverse shuffling operation $\mathcal{S}^{-1}$, the proposed shuffling equivalence regularization is the solution to  
    the inverse optimal transport problem.
\end{theorem}

\begin{proof}
    We provide a complete proof in \textcolor{Periwinkle} {\bf Appendix C.2}.
\end{proof}


\noindent \textbf{Implication.} Since the cost measure operates on instance embeddings instead of position coordinates, solving the inverse OT problem ensures that the learned cost measures can generate the transport plan $\mathcal{S}^{-1}$. Equivalently, this enforces the network to learn meaningful semantics between $\bm{\mathrm{X}}$ and $\bm{\mathrm{X}}'$ (\textit{i.e.,} $\mathcal{S}[\bm{\mathrm{X}}]$) such that the original instance ordering can be restored from their random arrangement. 

\begin{figure*}[ht]
\begin{minipage}{0.57\textwidth}
\captionof{table}{Main results on the CAMELYON16 dataset and TCGA-NSCLC dataset by using different feature extractors. Our method significantly outperforms all MIL-based competitors (see \textcolor{Periwinkle}{\bf Appendix E} for the statistical test).}
\label{tab:experiment_two_benchmark}
        \vspace{-0.1in}
		\resizebox{0.999\textwidth}{!}{
			\begin{tabular}{lccc|ccc}
				\toprule
                     & \multicolumn{3}{c}{CAMELYON16} & \multicolumn{3}{c}{TCGA-NSCLC} \\
                      \cmidrule(r){2-4} \cmidrule(r){5-7}
				  & Accuracy & F1 &  AUC & Accuracy & F1 &  AUC  \\
				\midrule
				& \multicolumn{6}{c}{\cellcolor{blue!20}\bf Swin-VIT ImageNet Pretrained } \\
                     ABMIL (\textit{ICML'18}) & $84.73_{0.85}$ & $83.20_{0.81}$ & $85.66_{1.76}$ & $91.07_{1.08}$ & $91.27_{1.23}$ & $95.88_{1.18}$ \\
DSMIL (\textit{CVPR'21}) & $84.42_{1.12}$ & $82.72_{1.15}$ & $86.69_{2.33}$ & $90.98_{1.49}$ & $90.97_{1.49}$ & $95.71_{0.18}$ \\
TransMIL (\textit{NeurIPS’21}) & $85.04_{1.70}$ & $83.72_{1.29}$ & $88.26_{0.88}$ & $89.73_{0.40}$ & $89.93_{0.62}$ & $95.66_{0.99}$ \\
MaxS (\textit{CVPR’22}) & $84.57_{1.22}$ & $78.87_{1.13}$ & $89.69_{1.25}$ & $87.33_{1.00}$ & $87.05_{1.31}$ & $93.09_{0.85}$ \\
AFS (\textit{CVPR’22}) & $79.61_{2.22}$ & $72.18_{0.95}$ & $83.88_{1.68}$ & $90.79_{1.52}$ & $90.36_{1.80}$ & $96.17_{0.89}$ \\
MaxMinS (\textit{CVPR’22}) & $83.80_{1.01}$ & $76.73_{1.29}$ & $86.78_{1.59}$ & $89.83_{0.87}$ & $89.44_{1.22}$ & $95.76_{0.57}$ \\
ILRA-MIL (\textit{ICLR’23}) & $84.96_{1.05}$ & $83.60_{0.86}$ & $87.76_{1.45}$ & $90.69_{1.13}$ & $90.68_{1.13}$ & $95.56_{0.97}$ \\
MHIM-MIL (\textit{ICCV'23}) & $86.24_{1.68}$ & $84.35_{2.15}$ & $86.12_{1.95}$ & $89.64_{1.66}$ & $89.61_{1.67}$ & $93.93_{0.84}$ \\
DGR-MIL (\textit{ECCV'24}) & $87.60_{2.39}$ & $86.47_{2.39}$ & $88.19_{1.73}$ & $90.88_{1.83}$ & $90.85_{1.84}$ & $95.81_{1.25}$ \\
AC-MIL (\textit{ECCV'24}) & $86.24_{1.01}$ & $84.94_{1.31}$ & $87.77_{1.61}$ & $90.50_{1.29}$ & $90.63_{1.23}$ & $95.61_{0.79}$ \\
\rowcolor{blue!8}\textbf{Ours [Trans.]} & $\textbf{89.53}_{1.40}$ & $\textbf{88.57}_{1.53}$ & $\textbf{92.17}_{0.49}$ & \underline{$\textit{92.32}_{1.25}$} & \underline{$\textit{92.31}_{1.26}$} & $\textbf{96.40}_{0.77}$ \\
\rowcolor{blue!8}\textbf{Ours [CNN]} & \underline{$\textit{88.11}_{0.58}$} & \underline{$\textit{87.11}_{0.63}$} & \underline{$\textit{91.80}_{0.26}$} & $\textbf{92.51}_{0.80}$ & $\textbf{92.49}_{0.81}$ &\underline{ $\textit{96.32}_{0.67}$}\\
                     
                 \hline
                    & \multicolumn{6}{c}{\cellcolor{blue!20}\bf ResNet-18 ImageNet Pretrained} \\
                    ABMIL (ICML'18) & $85.74_{0.99}$ & $84.21_{1.11}$ & $85.91_{1.53}$ & $88.10_{0.80}$ & $88.18_{0.82}$ & $93.88_{1.11}$ \\
DSMIL (\textit{CVPR'21}) & $84.19_{2.25}$ & $82.21_{2.82}$ & $84.84_{1.74}$ & $88.58_{1.02}$ & $88.61_{1.06}$ & $93.73_{0.87}$ \\
TransMIL (\textit{NeurIPS’21}) & $82.79_{1.89}$ & $76.63_{1.86}$ & $87.71_{1.84}$ & $84.65_{1.11}$ & $84.20_{0.90}$ & $90.71_{1.20}$ \\
MaxS (\textit{CVPR’22}) & $84.81_{2.09}$ & $83.62_{2.20}$ & $87.22_{1.78}$ & $88.39_{0.81}$ & $88.54_{1.00}$ & $93.43_{0.84}$ \\
AFS (\textit{CVPR’22}) & $81.94_{1.55}$ & $77.85_{1.45}$ & $89.23_{1.07}$ & $88.48_{0.81}$ & $88.27_{1.16}$ & $94.83_{0.92}$ \\
MaxMinS (\textit{CVPR’22}) & $82.02_{1.86}$ & $76.11_{0.88}$ & $88.04_{1.84}$ & $87.81_{0.86}$ & $87.5_{1.00}$ & $94.19_{0.95}$ \\
ILRA-MIL (\textit{ICLR’23}) & $87.08_{2.31}$ & $86.19_{2.56}$ & $89.30_{2.98}$ & $88.77_{0.98}$ & $88.8_{1.09}$ & $94.25_{0.68}$ \\
MHIM-MIL (\textit{ICCV'23}) & $86.05_{1.64}$ & $84.48_{1.82}$ & $86.17_{1.76}$ & $87.43_{1.37}$ & $87.4_{1.35}$ & $93.65_{0.62}$ \\
DGR-MIL (\textit{ECCV'24}) & $86.63_{0.85}$ & $85.25_{0.96}$ & $88.20_{1.30}$ & $87.43_{1.18}$ & $87.43_{1.14}$ & $93.88_{0.41}$ \\
AC-MIL (\textit{ECCV'24}) & $87.02_{1.49}$ & $85.55_{1.77}$ & $87.56_{2.37}$ & $88.58_{0.69}$ & $88.58_{0.69}$ & $94.31_{1.12}$ \\
\rowcolor{blue!8}\textbf{Ours [Trans.]} & \underline{$\textit{87.47}_{2.12}$} & \underline{$\textit{86.30}_{2.34}$} & $\underline{\textit{90.44}_{1.41}}$ & \underline{$\textit{88.96}_{0.97}$} & \underline{$\textit{89.02}_{0.98}$} & $\textbf{94.98}_{0.81}$ \\
\rowcolor{blue!8}\textbf{Ours [CNN]} & $\textbf{88.37}_{0.45}$ & $\textbf{87.16}_{0.36}$ & $\textbf{92.92}_{0.87}$ & $\textbf{90.40}_{0.98}$ & $\textbf{90.39}_{0.98}$ & \underline{$\textit{94.93}_{1.15}$}

\\

                \hline
                 & \multicolumn{6}{c}{\cellcolor{blue!20}\bf CTransPath Self-supervised Pretrained } \\
                    ABMIL (\textit{ICML'18}) & $94.80_{0.50}$ & $94.39_{0.55}$ & $96.50_{0.67}$ & $93.38_{1.10}$ & $93.36_{1.11}$ & $96.81_{0.63}$ \\
DSMIL (\textit{CVPR'21}) & $94.49_{0.64}$ & $94.08_{0.69}$ & $95.64_{0.56}$ & $94.24_{1.25}$ & $94.22_{1.26}$ & $97.85_{0.69}$ \\
TransMIL (\textit{NeurIPS’21}) & $94.42_{0.58}$ & $92.44_{0.68}$ & $97.34_{0.19}$ & $90.79_{0.72}$ & $90.39_{0.69}$ & $96.22_{0.79}$ \\
MaxS (\textit{CVPR’22}) & $94.96_{1.21}$ & $94.77_{1.15}$ & $97.33_{0.30}$ & $93.86_{0.95}$ & $93.85_{0.94}$ & $97.84_{0.52}$ \\
AFS (\textit{CVPR’22}) & $94.42_{0.68}$ & $92.42_{0.86}$ & $97.14_{0.27}$ & $93.28_{1.04}$ & $92.95_{1.17}$ & $97.81_{0.46}$ \\
MaxMinS (\textit{CVPR’22}) & $95.19_{0.47}$ & $93.38_{0.54}$ & $97.66_{0.44}$ & $93.66_{0.70}$ & $93.34_{0.74}$ & $97.78_{0.39}$ \\
ILRA-MIL (\textit{ICLR’23}) & $94.83_{1.77}$ & $94.45_{1.94}$ & $95.85_{1.00}$ & $93.57_{0.75}$ & $93.56_{0.75}$ & $97.44_{0.56}$ \\
MHIM-MIL (\textit{ICCV'23}) & $94.57_{0.55}$ & $94.16_{0.65}$ & $96.38_{0.61}$ & $93.95_{1.21}$ & $93.94_{1.21}$ & \underline{$\textit{97.87}_{0.53}$} \\
DGR-MIL (\textit{ECCV'24}) & $95.73_{1.16}$ & $95.41_{1.26}$ & $96.30_{0.47}$ & $94.53_{1.26}$ & $94.52_{1.27}$ & \underline{$\textit{97.87}_{0.50}$ }\\
AC-MIL (\textit{ECCV'24}) & $95.15_{0.64}$ & $94.88_{0.73}$ & $97.00_{0.69}$ & $94.72_{0.68}$ & $94.72_{0.68}$ & $97.76_{0.76}$ \\
\rowcolor{blue!8}\textbf{Ours [Trans.]} & $\textbf{96.64}_{0.37}$ & $\textbf{96.39}_{0.38}$ & \underline{$\textit{98.00}_{0.17}$} & $\textbf{95.20}_{1.23}$ & $\textbf{95.19}_{1.23}$ & $\textbf{97.99}_{0.67}$ \\
\rowcolor{blue!8}\textbf{Ours [CNN]} & \underline{$\textit{96.25}_{0.83}$} & \underline{$\textit{95.99}_{0.88}$} & $\textbf{98.10}_{0.31}$ & \underline{$\textit{95.11}_{0.83}$} & \underline{$\textit{95.10}_{0.84}$} & $97.55_{0.77}$\\ 
                     
				\bottomrule
 			\end{tabular}}
\end{minipage}
\hspace{0.5cm}
\begin{minipage}{0.37\textwidth}
    \includegraphics[width=0.95\textwidth]{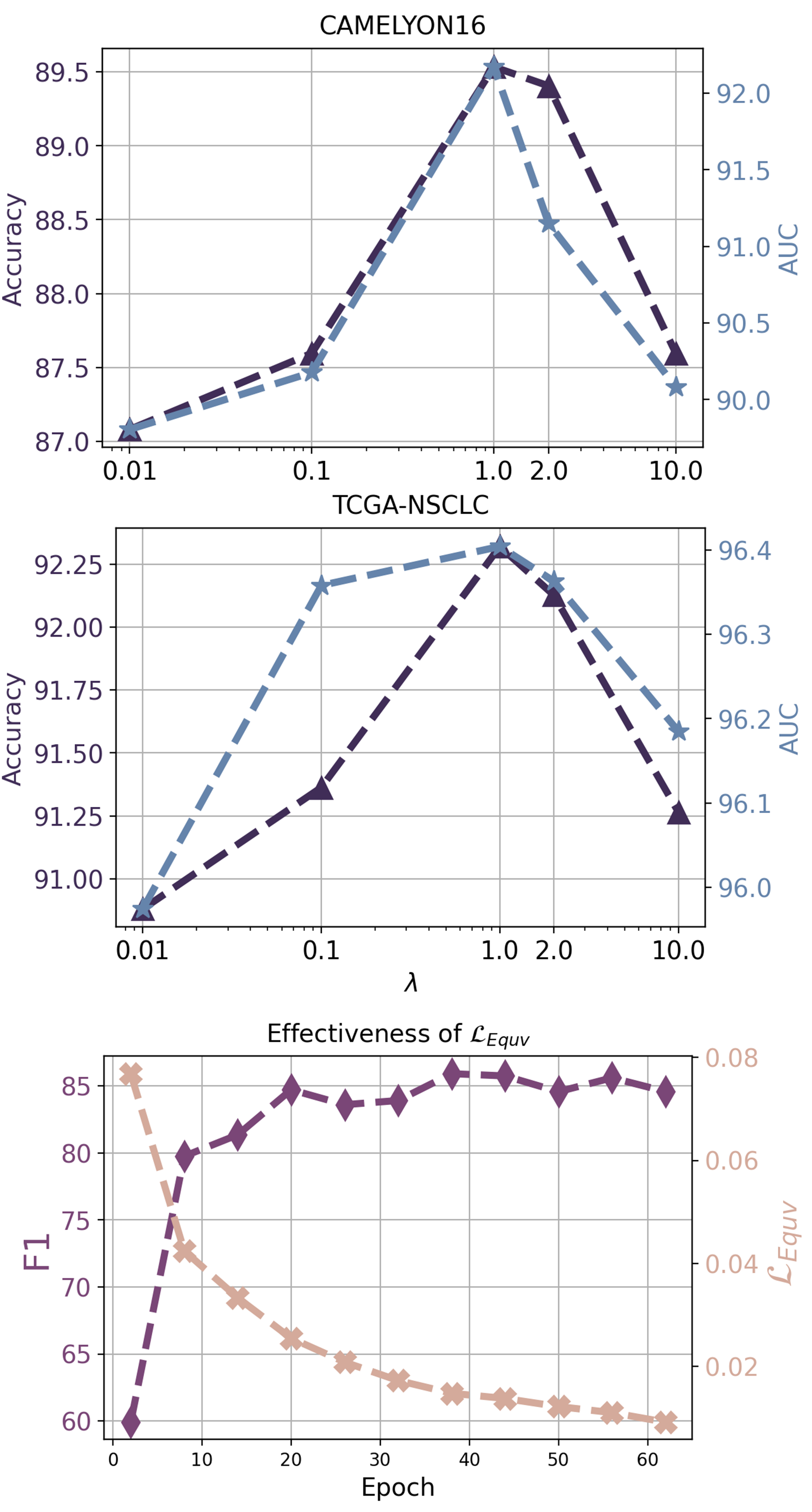} 
    
    \caption{ \textbf{Top:} The effectiveness of $\lambda$ on CAMELYON16 datasets, \textbf{Middle:} \textbf the effectiveness of $\lambda$ on TCGA-NSCLC datasets, and \textbf{Bottom:} Training dynamics for $\mathcal{L}_{\text{Equv}}$ and performance.  }
    \label{fig:aba}
\end{minipage}
\end{figure*}

\section{Experiments and Results}
In this section, we rigorously validate our methods on two main tasks for WSIs: (i) \textit{WSI classification} and (ii) \textit{WSI survival prediction}. We compare the proposed method with recent state-of-the-art MIL methods, including: ABMIL~\cite{ilse2018attention},  DSMIL~\cite{li2021dual}, Trans-MIL~\cite{shao2021transmil}, DTFD-MIL~\cite{zhang2022dtfd} with three distillation strategies, ILRA-MIL~\cite{lowrankmil}, MHIM-MIL \cite{tang2023multiple}, DGR-MIL \cite{zhu2025dgr}, and AC-MIL \cite{zhang2023attention}. See our discussion and classification of these methods in Sec.~\ref{sec:2}. For the details of the implementation, please refer to \textcolor{Periwinkle} {\bf Appendix D}.


\subsection{Results on WSI Classification}
\noindent\textbf{Setup.} We validate our method on the CAMELYON16 dataset and TCGA-NSCLC dataset for WSI classification. The CAMELYON16 dataset is a publicly available collection of WSIs designed to detect metastatic breast cancer in lymph node tissue, while the TCGA-NSCLC dataset is a public dataset for classifying two subtypes of lung cancer (lung squamous cell carcinoma and adenocarcinoma). Following the protocols outlined in~\cite{clam-sb}, each slide is processed into around $11,000$ and $13,000$ patches at \texttt{20$\times$} magnification on average for CAMELYON16 and TCGA-NSCLC
(see \textcolor{Periwinkle} {\bf Appendix D} for more details about the dataset pre-processing). For a more rigorous comparison, we conduct experiments on instance feature embeddings extracted by three feature extractors. Specifically, we select three different feature extractors that vary in architecture and pretraining strategies: \textbf{(i)} ResNet-18 pre-trained on ImageNet~\cite{resnet}; \textbf{(ii)} a Swin Vision Transformer (Swin-ViT) pre-trained on ImageNet~\cite{swimtransformer}; and \textbf{(iii)} CTransPath~\cite{wang2021transpath} pre-trained on large-scale histopathological datasets. 
For experiments on CAMELYON16, we use the official train/test split and report results averaged over five runs. Following the setup outlined in~\cite{zhang2022dtfd}, we perform standard 4-fold cross-validation on TCGA-NSCLC and present the mean ($\pm$ standard deviation) of the results. We evaluate the performance with \textit{accuracy}, \textit{F1 score}, and \textit{AUC}.

\noindent\textbf{Results.} The main results for WSI classification are shown in Table \ref{tab:experiment_two_benchmark}. The proposed method, implemented by two different network architectures (\textit{i.e.,} CNN and transformer), consistently and significantly outperforms the MIL-based methods using features extracted by different means across both datasets. Specifically, our transformer-based variant surpasses the best MIL model by an average of 1.84\% and 0.72\% across three metrics on CAMELYON16 and TCGA-NSCLC, respectively, using feature pre-trained on ImageNet (\textit{i.e.,} Swin-ViT and ResNet-18 pre-trained on ImageNet in Table \ref{tab:experiment_two_benchmark}). Similarly, our CNN-based variant surpasses the best MIL model by an average of 2.03\% and 1.27\% across all three metrics on CAMELYON16 and TCGA-NSCLC. Although the improvement is relatively smaller, the two variants of our method outperform the best MIL model by an average of 1.13\% and 0.30\% across all metrics on CAMELYON16 and TCGA-NSCLC, respectively. This is because the self-supervised pretraining already produces high-quality instance feature representations that effectively capture their similarities.

\begin{table}[]
\centering
\caption{Comparison of the performance by different methods for survival prediction in terms of \textit{C-Index} using \textbf{UNI}~\cite{chen2024uni} features. }
\label{tab:survival prediction}
\vspace{-0.1in}
\resizebox{0.85\linewidth}{!}{%
\begin{tabular}{lcc} \toprule
 & TCGA-LUAD & TCGA-BRCA \\ \midrule
ABMIL \textit{(ICML'18)} & $0.659_{0.022}$ & $0.728_{0.050}$ \\
DSMIL \textit{(CVPR'21)} & $0.656_{0.029}$ & $0.715_{0.052}$ \\
TransMIL\textit{ (NeurIPS’21)} & $0.609_{0.032}$ & $0.721_{0.023}$ \\
DTFD-MIL    \textit{(CVPR’22)} & $0.656_{0.046}$ & $0.733_{0.022}$ \\
ILRA-MIL \textit{(ICLR’23)} & $0.595_{0.032}$ & $0.718_{0.026}$ \\
MHIM-MIL \textit{(ICCV'23)} & $0.650_{0.020}$ & $0.736_{0.031}$ \\
DGR-MIL \textit{(ECCV'24)} & $0.620_{0.013}$ & $0.711_{0.038}$ \\
AC-MIL \textit{(ECCV'24)} & $0.656_{0.017}$ & $0.659_{0.062}$ \\
\rowcolor{blue!8}\textbf{Ours [Trans.]} & \underline{$\textit{0.677}_{0.028}$} & \underline{$\textit{0.752}_{0.017}$} \\
\rowcolor{blue!8}\textbf{Ours [CNN]} & {$\textbf{0.683}_{0.030}$} & {$\textbf{0.757}_{0.018}$} \\ \bottomrule
\end{tabular}%
}
\vspace{-0.1cm}
\end{table}

\subsection{Results on WSI Survival Prediction}
The survival prediction aims to estimate the risk of a certain event of interest (\textit{e.g.,} death). The data is typically structured as a triplet ($\bm{\mathrm{X}}_i$, $t_i$, $e_i$), where $\bm{\mathrm{X}}_i$ is the WSI instance features, $t_i$ is the recorded time, and $e_i$ indicates the if an event of interest happens ($e_i=1$ if the event happens; $e_i =0$ for censored observation\footnote{The event of interest is not observed at the end of study.}). 

\noindent\textbf{Setup.} 
We validate the proposed method on two widely used datasets for survival prediction: TCGA-LUAD and TCGA-BRCA. Following the experimental setup in~\cite{jaume2024modeling}, we conduct a five-fold cross-validation. Following the method outlined in~\cite{chen2021whole,jaume2024modeling}, we consider a discrete-time survival prediction setting, which predicts if the event happens at each time interval $(t_{i, j-1}, t_{i, j})$. This formulates a classification problem under censorship, where a negative log-likelihood survival loss \cite{zadeh2020bias} is adopted to optimize the model. We still consider previous MIL models as the baselines, since we only focus on WSI features. Please refer to \textcolor{Periwinkle} {\bf Appendix D} for additional details on the implementation. 
In line with prior studies \cite{jaume2024modeling,song2024multimodal}, performance is evaluated by the concordance index (\textit{C-Index}), which assesses the ordering performance among all the uncensored and censored pairs, with values ranging from 0 to 1. 

\noindent\textbf{Results.} The results of survival prediction are shown in Table \ref{tab:survival prediction}. The two variants of our method consistently outperform the MIL-based methods across both datasets. In particular, the transformer-based and CNN-based variants outperform the second-best method by an average of 3.16\% and 3.97\%, respectively, across both datasets. We attribute the superiority of the CNN-based variant to its ability to accurately capture local structures that are closely tied to changes in the survival rate. We would also like to point out that even with the same network architecture, our transformer-based variant surpasses TransMIL by an average of 5.61\% across two datasets in terms of C-Index. This further confirms the superiority of the proposed shuffling equivalence regularization over the permutation invariance used in MIL.

\subsection{Ablation Studies}\label{sec:ablation}
Unless specified otherwise, experiments in Sec.~\ref{sec:ablation} and Sec.~\ref{sec:analysis} are conducted using features extracted by Swin-ViT pretrained on ImageNet. In this section, we provide ablation studies on the proposed shuffling equivalence regularization mechanism. 

\noindent\textbf{Effectiveness.} We conduct the ablation on the effectiveness of the proposed shuffling equivalence regularization on both CAMELYON16 and TCGA-NSCLC. As shown in Table~\ref{tab:ablation}, the introduction of the proposed shuffling equivalence regularization into both transformer and CNN leads to substantial performance gain across both datasets. Specifically, our two variants can achieve a performance gain of at least 2\% across all metrics on CAMELYON16. A similar trend is observed on TCGA-NSCLC, where there is a performance gain of 1.72\%, 1.39\%, and 0.6\% in accuracy, F1 score, and AUC, respectively. We also confirm that the minimization of the shuffling equivalence regularization loss leads to performance improvement (see evidence in training dynamics shown in Fig.~\ref{fig:aba} (\textbf{bottom})).

\noindent\textbf{Regularization Strength.} We investigate the effectiveness of the regularization strength, \textit{i.e.,} $\lambda$ presented in Eq. \ref{eq:weight}. The corresponding results are illustrated in Fig.~\ref{fig:aba} (\textbf{top} and \textbf{middle}), where the optimal $\lambda$ is set to 1 for both datasets. Though there is some fluctuation in the performance under different regularization strengths, we do not observe significant changes in performance.  However, a too small or large regularization could degrade the performance.

\begin{table}[]
\centering
\caption{The ablation studies on the shuffling equivalence regularization  ($\mathcal{L}_{\text{Equv}}$) on CAMELYON16 and TCGA-NSCLC. Please refer to \textcolor{Periwinkle} {\bf Appendix B.4} for ablation using \textbf{CtransPath} features.}
\label{tab:ablation}
\vspace{-0.1in}
\resizebox{0.74\linewidth}{!}{%
\begin{tabular}{cccccc}\toprule
Dataset & Network & $\mathcal{L}_{\text{Equv}}$ & \multicolumn{1}{c}{Accuracy} & \multicolumn{1}{c}{F1} & \multicolumn{1}{c}{AUC} \\ \midrule
\multirow{6}{*}{\rotatebox{0}{CAMELYON16}} & \multirow{3}{*}{\textbf{Trans.}} & \ding{55} & 86.51 & 85.17 & 88.89 \\
 &  & \ding{51} & 89.53 & 88.57 & 92.17 \\
 &  & \cellcolor{blue!8}$\Delta$ &\cellcolor{blue!8} \textbf{+3.02} & \cellcolor{blue!8}\textbf{+3.40} & \cellcolor{blue!8}\textbf{+3.28} \\  \cline{2-6}
 & \multirow{3}{*}{\textbf{CNN}} & \ding{55} & 85.58 & 84.12 & 89.38 \\
 &  & \ding{51} & 88.11 & 87.11 & 91.80 \\
 &  &\cellcolor{blue!8} $\Delta$ &\cellcolor{blue!8}\textbf{ +2.53} &\cellcolor{blue!8} \textbf{+2.99} & \cellcolor{blue!8}\textbf{ +2.43} \\ \midrule

\multirow{6}{*}{\rotatebox{0}{TCGA-NSCLC}} & \multirow{3}{*}{\textbf{Trans.}} & \ding{55} & 90.60 & 90.92 & 95.80 \\
 &  & \ding{51} & 92.32 & 92.31 & 96.40 \\
 &  &\cellcolor{blue!8} $\Delta$ & \cellcolor{blue!8}\textbf{+1.72} & \cellcolor{blue!8}\textbf{+1.39} &\cellcolor{blue!8}\textbf{+0.60} \\  \cline{2-6}
 & \multirow{3}{*}{\textbf{CNN}} & \ding{55} & 90.21 & 90.20 & 95.65 \\ 
 &  & \ding{51} & 92.51 & 92.49 & 96.32 \\
 &  & \cellcolor{blue!8}$\Delta$ & \cellcolor{blue!8}\textbf{+2.30} & \cellcolor{blue!8}\textbf{+2.29} & \cellcolor{blue!8}\textbf{+0.67} \\ \bottomrule
\end{tabular}%
}
\vspace{-0.2in}
\end{table}

\begin{figure*}[!t]
    \centering
    \includegraphics[width=1.0\linewidth]{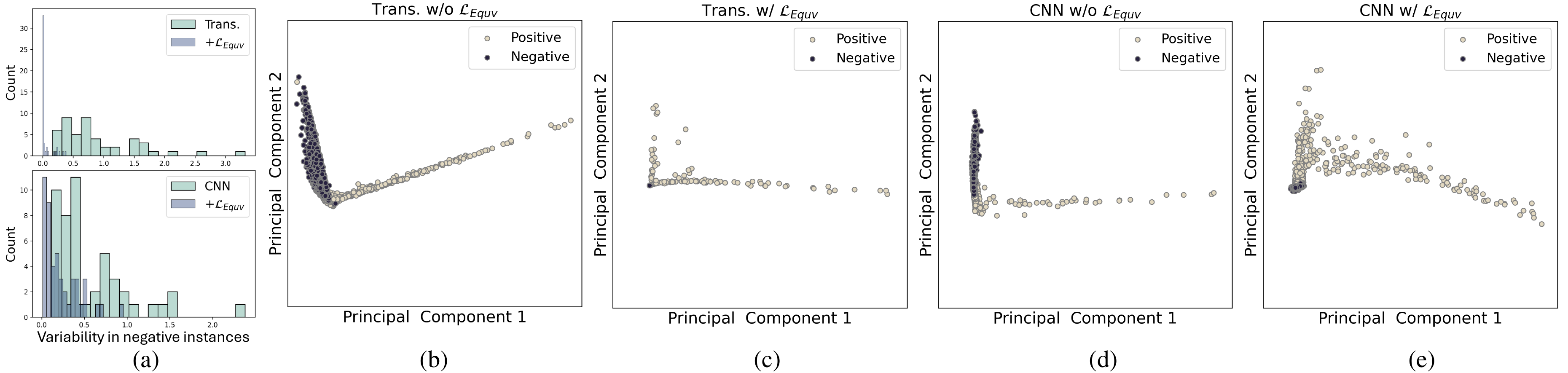}
    \vspace{-0.2in}
    \caption{(a) The histogram of variability of the negative instance feature representations from different WSIs on CAMELYON16 (please refer to~\citep[][Eq.~4]{papyan2020prevalence} for details). (b-e) The learned instance-level feature representations by different variants of our method are visualized by projecting them into two dimensions using PCA. The proposed method with equivalence regularization shows small variability, which corresponds to the collapse of variability when the learned representations converge to a compact and uniform structure without redundancy.}
    \label{fig:inst_neural_collapse}
\end{figure*}

\begin{figure}
    \centering
    \includegraphics[width=0.99\linewidth]{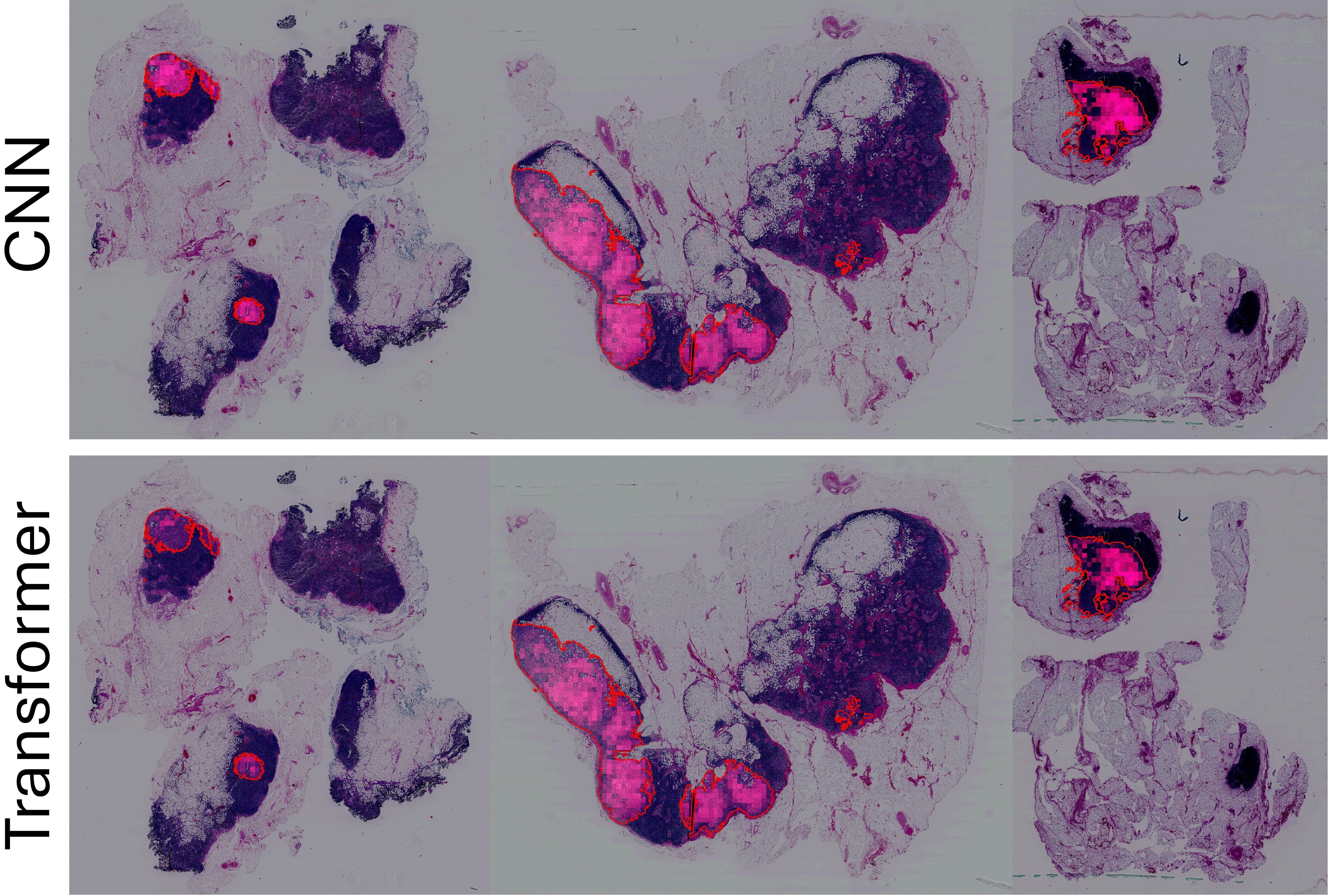}
    \caption{CAM produced by our CNN (\textbf{top}) and Transformer (\textbf{bottom}) variants on CAMELYON16 (Swin-ViT features).}
    \label{fig:cam}
    \vspace{-0.2in}
\end{figure}

\subsection{Analysis and Discussion}\label{sec:analysis}
\vspace{-0.06in}
In this section, we offer some insights into why the proposed method is effective.

\noindent\textbf{Instance-level Representations.} We compare the learned instance-level feature representations from our method and the corresponding baseline methods without our equivalence regularization on CAMELYON16. For visualization, we reduce the dimensionality to two using principal component analysis (PCA). Interestingly, we observe that the negative instance representations learned by the proposed method show a small variability compared to standard MIL-based methods (see Fig.~\ref{fig:inst_neural_collapse}(a)). A similar trend is observed in the visualization of the learned negative instance feature representations (see Fig.~\ref{fig:inst_neural_collapse}(c) and (e)), where the negative instance representations are collapsed to a single point or form a tightly clustered group with small variation. However, this phenomenon disappears when our shuffling equivalence regularization is removed (Fig.~\ref{fig:inst_neural_collapse}(b) and (d)).

\noindent\textbf{Negative Instance Collapse.} The small variability of the learned instance representations aligns with the phenomena of \textit{collapse of variability} discussed in~\cite{papyan2020prevalence,zhu2021geometric,zhou2022all}, where the optimal solution of the classification task results in a class-wise representation collapse. This is particularly advantageous for WSI analysis, as a simple decision boundary can effectively distinguish between positive and negative instances with homogeneity. However, the same phenomenon is not observed among the positive instances. We hypothesize that this occurs because negative instances make up most of the WSIs, and collapse tends to occur in populations with more samples (\textit{i.e.,} negative instances in our case). 
The diversity in positive instance representations indicates a sensitivity to the misclassification of positive instances (\textit{i.e.,} false negatives), which is particularly important in most WSI applications. 
The collapse of negative instances provides a meaningful explanation for the superior performance of our method.


\noindent\textbf{CNN vs. Transformer.} Overall, our CNN variant shows slightly better performance than our transformer variant in both WSI classification and survival prediction (see Tables~\ref{tab:experiment_two_benchmark} and~\ref{tab:survival prediction}). Similarly, the CNN variant can localize the region of interest better than the transformer variant (Fig.~\ref{fig:cam}). We hypothesize that this is a unique advantage of CNNs, attributed to their inherent inductive biases, \textit{e.g.,} locality.

\section{Conclusion and Future Work}
In this work, we challenge the necessity of standard permutation-invariant MIL for WSI analysis by highlighting the importance of semantic information among instances. In light of this, we introduce an alternative to MIL for WSI analysis, which explicitly learns the semantics between instances by solving \textit{instance jigsaw puzzles}. Our novel Siamese network solution to the instance jigsaw puzzles is both theoretically sound and empirically effective, outperforming the previous MIL-based methods in WSI classification and survival prediction. 
Despite the slightly increased computational burden, our approach paves the way for various network structures (\textit{e.g.,} CNNs) for WSI analysis, allowing the exploitation of their unique properties. This expands the solution space for WSI analysis and offers opportunities for exploration in future work.

{
    \small
    \bibliographystyle{ieeenat_fullname}
    \bibliography{main}
}


\appendix
\renewcommand{\thefigure}{S\arabic{figure}}
\renewcommand{\thetable}{S\arabic{table}}

\section{Proofs in Section 3}

\subsection{Proof of Proposition 1}\label{sec:ap:1}
Without loss of generality, we define any random permutation/shuflling of a bag of instances as $\bm{\mathrm{X}}_{\sigma} = \mathcal{S}[\bm{\mathrm{X}}] = \{\bm{\mathrm{x}}_{\sigma(1)}, \bm{\mathrm{x}}_{\sigma(2)},\cdots, \bm{\mathrm{x}}_{\sigma(n)}\} \in \mathbb{R}^{n \times d}$, where $\sigma$ is a permutation of indices $\{1, 2, \cdots, n\}$. If we further define a permutation matrix $\bm{\mathbf{P}}_\sigma \in \mathbb{R}^{n \times n}$, we have 
\begin{align*}
    \bm{\mathrm{X}}_{\sigma} = \bm{\mathbf{P}}_\sigma \bm{\mathbf{X}},
\end{align*}
where each row and each column in $\mathbf{P}_\sigma$ has exactly one element equal to 1, with the other elements being zero. $(\mathbf{P}_\sigma)_{ij}=1$ implies the $i$-th instance of $\bm{\mathrm{X}}_{\sigma}$ from the $j$-th instance of $\bm{\mathbf{X}}$.
It is easy to verify that $\bm{\mathbf{P}}_\sigma$ is an orthonormal matrix such that $\bm{\mathbf{P}}_\sigma^\top \bm{\mathbf{P}}_\sigma = \bm{\mathbf{P}}_\sigma \bm{\mathbf{P}}_\sigma^\top = \bm{\mathrm{I}}$. This is because $\bm{\mathbf{P}}_\sigma^\top = \mathbf{P}_\sigma^{-1}$ denotes the inverse process of permutation, which should restore the original order of instances.

We start with introducing the following lemma and corollary to support the proof of \textbf{Proposition 1}.

\begin{lemma}\label{lemma:s1}
For any element-wise activation function $\operatorname{act}(\cdot)$, the following permutation equivalence holds:
\begin{align*}
    \operatorname{act}(\mathbf{P}_\sigma\mathbf{X}) = \mathbf{P}_\sigma \operatorname{act}(\mathbf{X}).
\end{align*}
\end{lemma}
\begin{proof}
    Since the activation function is applied element-wise to the permuted input, we have 
    \begin{align*}
        \operatorname{act}(\mathbf{P}_\sigma\mathbf{X}) &= [\operatorname{act}(\mathbf{x}_{\sigma(1)}), \operatorname{act}(\mathbf{x}_{\sigma(2)}), \cdots, \operatorname{act}(\mathbf{x}_{\sigma(n)})] \\
        &= [\operatorname{act}(\mathbf{x_1})_{\sigma(1)}, \operatorname{act}(\mathbf{x}_2)_{\sigma(2)}, \cdots, \operatorname{act}(\mathbf{x}_n)_{\sigma(n)}]\\
        &= \mathbf{P}_\sigma \operatorname{act}(\mathbf{X}).
    \end{align*}
This completes the proof.
\end{proof}

\begin{corollary}\label{col:s1}
Lemma~\ref{lemma:s1} can also be extended to the following form:
\begin{align*}
    \operatorname{act}(\mathbf{X}^\top\mathbf{P}_\sigma^\top) &=  \operatorname{act}(\mathbf{P}_\sigma \mathbf{X})^\top \\
    &= \left(\mathbf{P}_\sigma \operatorname{act}(\mathbf{X}) \right)^\top\\
    &=\operatorname{act}(\mathbf{X}^\top)\mathbf{P}_\sigma^\top. 
\end{align*}
\end{corollary}

\subsubsection{ABMIL}
The standard attention pooling ($\operatorname{Attn-Pool}(\cdot)$) in ABMIL without any positional encoding can be described as 
\begin{align*}
    \operatorname{Attn-Pool}(\bm{\mathrm{X}}) = \underbrace{\operatorname{softmax}\left(\bm{\mathrm{W}}^\top \operatorname{tanh}(\bm{\mathrm{V}}  \bm{\mathrm{X}}^\top) \right)}_{\mathbb{R}^{1\times n}} \bm{\mathrm{X}},
\end{align*}
where $\bm{\mathrm{V}} \in \mathbb{R}^{L \times d}$ and $\bm{\mathrm{W}} \in \mathbb{R}^{L \times 1}$ are learnable weight matrices.  We want to prove that
\begin{align*}
    \operatorname{Attn-Pool}(\bm{\mathrm{X}}_\sigma) = \operatorname{Attn-Pool}(\bm{\mathrm{X}}).
\end{align*}
Applying Corollary~\ref{col:s1}, the above attention pooling for a permuted bag of instances can be described as 
\begin{align*}
    &\operatorname{Attn-Pool}(\bm{\mathrm{X}}_\sigma)\\ \nonumber
    =& \operatorname{softmax}\left(\bm{\mathrm{W}}^\top \operatorname{tanh}(\bm{\mathrm{V}}  (\bm{\mathrm{P}}_\sigma \bm{\mathrm{X}})^\top) \right) (\bm{\mathrm{P}}_\sigma \bm{\mathrm{X}}) \\ \nonumber
    =& \operatorname{softmax}\left(\bm{\mathrm{W}}^\top \operatorname{tanh}(\bm{\mathrm{V}}  \bm{\mathrm{X}}^\top \bm{\mathrm{P}}_\sigma^\top) \right)(\bm{\mathrm{P}}_\sigma \bm{\mathrm{X}}) \\ \nonumber
    =& \operatorname{softmax}\left(\bm{\mathrm{W}}^\top \operatorname{tanh}(\bm{\mathrm{V}}  \bm{\mathrm{X}}^\top) \right)(\bm{\mathrm{P}}_\sigma^\top\bm{\mathrm{P}}_\sigma \bm{\mathrm{X}}) \\ \nonumber
    =&  \operatorname{Attn-Pool}(\bm{\mathrm{X}}).
\end{align*}
This completes the proof that attention-based pooling in ABMIL is permutation-invariant. 

\subsubsection{TransMIL}
The standard self-attention without positional encoding can be described as follows:
\begin{align*}
    & \operatorname{Self-Attn}(\bm{\mathrm{X}}) = \operatorname{softmax}\left( \frac{\bm{\mathrm{Q}}\bm{\mathrm{K}}^{\top}}{\sqrt{d_k}} \right)\bm{\mathrm{V}}  \\
    & \bm{\mathrm{Q}} = \bm{\mathrm{X}} \bm{\mathrm{W}}^{Q}, \bm{\mathrm{K}} = \bm{\mathrm{X}} \bm{\mathrm{W}}^{K}, \bm{\mathrm{V}} = \bm{\mathrm{X}} \bm{\mathrm{W}}^{V},
\end{align*}
where $\bm{\mathrm{W}}^{Q}$, $\bm{\mathrm{W}}^{K}$, $\bm{\mathrm{W}}^{V}$ $\in \mathbb{R}^{d \times d_k}$ are learnable weight matrices. 
Applying Lemma~\ref{lemma:s1} and Corollary~\ref{col:s1}, the self-attention evaluated on a permuted bag of instances can be described as 
\begin{align*}
    &\operatorname{Self-Attn}(\bm{\mathrm{X}}_\sigma) \\ \nonumber
    =& \operatorname{softmax}\left( \frac{(\bm{\mathrm{P}}_\sigma\bm{\mathrm{Q}})(\bm{\mathrm{P}}_\sigma\bm{\mathrm{K}})^{\top}}{\sqrt{d_k}} \right) (\bm{\mathrm{P}}_\sigma\bm{\mathrm{V}})  \\ \nonumber
    =& \operatorname{softmax}\left( \frac{(\bm{\mathrm{P}}_\sigma\bm{\mathrm{Q}})\bm{\mathrm{K}}^{\top}}{\sqrt{d_k}} \right) (\bm{\mathrm{P}}_\sigma^\top \bm{\mathrm{P}}_\sigma\bm{\mathrm{V}}) \\ \nonumber
    =& \bm{\mathrm{P}}_\sigma \operatorname{Self-Attn}(\bm{\mathrm{X}}).
\end{align*}
The above result can be easily extended to a transformer layer with self-attention blocks.

\noindent \textbf{Final Pooling.} If we consider a global average pooling or max pooling after the final transformer layer, we have 
\begin{align*}
    \operatorname{avgpool}\left(\operatorname{Self-Attn}(\bm{\mathrm{X}}_\sigma)\right) &=  \operatorname{avgpool}\left(\operatorname{Self-Attn}(\bm{\mathrm{X}}) \right) \\ 
    \operatorname{maxpool}\left(\operatorname{Self-Attn}(\bm{\mathrm{X}}_\sigma)\right) &=  \operatorname{maxpool}\left(\operatorname{Self-Attn}(\bm{\mathrm{X}}) \right) 
\end{align*}
Although the self-attention is permutation-equivariant instead of permutation-invariant, applying the permutation-invariant global average pooling or max pooling on top of it ensures permutation invariance ~\citep[][Sec. 3]{zaheer2017deep}. 

\noindent \textbf{Class Token.} In the case of adding a class token $\bm{\mathrm{x}}_{cls}$ (instead of final pooling) to the instances as in TransMIL~\cite{shao2021transmil}, the permutation-invariance of TransMIL is trivial to verify. This is because the attention between the output of the class token is invariant to the permutation of the input tokens. 

\subsubsection{General Attention-based Pooling Mechanisms}
The above verification of permutation invariance can also be extended to other attention-based pooling mechanisms, which typically involves a dot product between input instances, \textit{i.e.,} $\bm{\mathbf{X}}^\top \bm{\mathbf{X}}$.
\begin{lemma}\label{lemma:s3}
For the orthonormal matrix $\bm{\mathrm{P}}_\sigma$, we have the following permutation-invariant property:
\begin{align*}
    \bm{\mathrm{X}}_\sigma^\top  \bm{\mathrm{X}}_\sigma &= (\bm{\mathbf{P}}_\sigma \bm{\mathbf{X}})^\top (\bm{\mathbf{P}}_\sigma \bm{\mathbf{X}}) \\
    &= \bm{\mathbf{X}}^\top \bm{\mathbf{P}}_\sigma^\top \bm{\mathbf{P}}_\sigma\bm{\mathbf{X}} \\
    &= \bm{\mathbf{X}}^\top \bm{\mathbf{X}}.
\end{align*}
\end{lemma}
Lemma~\ref{lemma:s3} in conjunction with Lemma~\ref{lemma:s1} and Corollary~\ref{col:s1} are the key for proving the permutation-invariance in the case of ABMIL and TransMIL. The same principal can be generalized to verify the permutation invariance in more general attention-based pooling mechanisms. However, this is beyond the score of this paper, we leave it to the future exploration. 

\subsubsection{Permutation-variance with Positional Encoding}
We define a bag of instances with positional encoding as
\begin{align*}
    \mathbf{X}_{\text{PE}} = \mathbf{X} + \mathbf{PE},
\end{align*}
where $\mathbf{PE}=[\mathrm{PE}_1, \mathrm{PE}_2, \cdots, \mathrm{PE}_n]$, with $\mathrm{PE}_n$ corresponding to the positional encoding for the $n$-th instance in a bag.
Likewise, the permuted $n$ instances with positional encoding is denoted as
\begin{align*}
    (\bm{\mathrm{X}}_{\text{PE}})_\sigma = \bm{\mathrm{X}}_\sigma + \mathbf{PE}.
\end{align*}
In the case of any permutation-invariant MIL (denoted as $\operatorname{MIL}(\cdot)$), we have
\begin{align*}
    \operatorname{MIL}((\bm{\mathrm{X}}_{\text{PE}})_\sigma)&=\operatorname{MIL}(\mathbf{P}_\sigma^\top (\bm{\mathrm{X}}_{\text{PE}})_\sigma) \\ \nonumber
    &= \operatorname{MIL}(\mathbf{P}_\sigma^\top (\bm{\mathrm{X}}_\sigma + \mathbf{PE})) \\ \nonumber
    &= \operatorname{MIL}(\mathbf{P}_\sigma^\top (\mathbf{P}_\sigma\bm{\mathrm{X}} + \mathbf{PE})) \\ \nonumber
    &= \operatorname{MIL}(\mathbf{X}+\bm{\mathbf{P}}_\sigma^\top \mathbf{PE} ).
\end{align*}
For any non-trivial $\mathbf{PE} \neq \mathbf{I}$, 
\begin{align*}
    \mathbf{X}+\bm{\mathbf{P}}_\sigma^\top \mathbf{PE} = \mathbf{X}+\mathbf{PE},
\end{align*}
\textit{if and only if} $\mathbf{P}_\sigma = \mathbf{I}$. This immediately suggests that there is no non-trivial permutation $\mathbf{P}_\sigma$ and $\mathbf{PE}$ to ensure the permutation invariance in MIL when adding positional encoding. 
Hence, we conclude that models with positional encoding are not generally permutation-invariant.


\subsection{Proof of Theorem 2}

\begin{theorem}
Incorporating positional information can lower the classification-error upper bound. This is because
\begin{align*}
H(\boldsymbol{\mathrm{Y}}|\boldsymbol{\mathrm{X}}) \geq H(\boldsymbol{\mathrm{Y}}|\boldsymbol{\mathrm{X}}, \boldsymbol{\mathrm{P}}), 
\end{align*}
where $\boldsymbol{\mathrm{P}}= \{\boldsymbol{\mathrm{p}}_1, \boldsymbol{\mathrm{p}}_2, \cdots, \boldsymbol{\mathrm{p}}_n\}$ denotes the positional coordinates associated with each instance on the raw WSIs.
\end{theorem}

To prove \textbf{Theorem 2}, we first introduce the upper bound of the classification error in Lemma~\ref{lemma:1}.
\begin{lemma}\label{lemma:1}
  (\cite{hellman1970probability})  The Bayesian classification error 
$
P_e = \int P_X(x) \left[ 1 - \max_{\mathbf{Y} }P(\mathbf{Y}|\mathbf{X}) \right] \text{d}\mathbf{X},
$ is bounded by:
\begin{align*}
P_e \leq \frac{1}{2} H(\mathbf{Y}|\mathbf{X}),
\end{align*}
where $P(\mathbf{Y}|\mathbf{X})$ is the posterior probability of the class label $\mathbf{Y}$ given a bag $\mathbf{X}$. $H(\cdot)$ denotes the entropy. 
\end{lemma}
\begin{remark}
    According to Lemma~\ref{lemma:1}, \textbf{without} incorporating the position information, the upper bound of the classification error is directly presented as $ H(\boldsymbol{\mathrm{Y}}|\boldsymbol{\mathrm{X}})$. When incorporating the position information $\boldsymbol{\mathrm{P}}$, the upper bound is presented as $H(\boldsymbol{\mathrm{Y}}|\boldsymbol{\mathrm{X}},  \boldsymbol{\mathrm{P}})$.
\end{remark} 
\noindent Below, we begin to prove \textbf{Theorem 2} by showing that $H(\boldsymbol{\mathrm{Y}}|\boldsymbol{\mathrm{X}},  \boldsymbol{\mathrm{P}})$ is a tighter error bound than $ H(\boldsymbol{\mathrm{Y}}|\boldsymbol{\mathrm{X}})$, \textit{i.e.,} $H(\boldsymbol{\mathrm{Y}}|\boldsymbol{\mathrm{X}},  \boldsymbol{\mathrm{P}}) \leq H(\boldsymbol{\mathrm{Y}}|\boldsymbol{\mathrm{X}})$.
\begin{proof}
According to the definition of entropy \cite{cover1999elements}, $H(\boldsymbol{\mathrm{Y}}|\boldsymbol{\mathrm{X}})$ can be presented as
\begin{align*}
    H(\boldsymbol{\mathrm{Y}}|\boldsymbol{\mathrm{X}}) = -\mathbb{E}_{{\mathcal{X}}, {\mathcal{Y}}} \left[ \log P(\boldsymbol{\mathrm{y}}|\boldsymbol{\mathrm{x}}) \right].
\end{align*}
Likewise, $H(\boldsymbol{\mathrm{y}}|\boldsymbol{\mathrm{x}}, \boldsymbol{\mathrm{p}})$ is presented as
\begin{align*}
    H(\boldsymbol{\mathrm{Y}}|\boldsymbol{\mathrm{X}}, \boldsymbol{\mathrm{P}}) = -\mathbb{E}_{{\mathcal{X}}, {\mathcal{Y}},{\mathcal{P}}} \left[ \log P(\boldsymbol{\mathrm{y}}|\boldsymbol{\mathrm{x}}, \boldsymbol{\mathrm{p}}) \right].
\end{align*}
By their definition,
\begin{align}\label{eq:thm2-eq1}
    &H(\boldsymbol{\mathrm{Y}}|\boldsymbol{\mathrm{X}})- H(\boldsymbol{\mathrm{Y}}|\boldsymbol{\mathrm{X}}, \boldsymbol{\mathrm{P}}) \\ \nonumber
    =&  -\mathbb{E}_{{\mathcal{X}}, {\mathcal{Y}}} \left[ \log P(\boldsymbol{\mathrm{y}}|\boldsymbol{\mathrm{x}}) \right]+\mathbb{E}_{{\mathcal{X}}, {\mathcal{Y}},{\mathcal{P}}} \left[ \log P(\boldsymbol{\mathrm{y}}|\boldsymbol{\mathrm{x}}, \boldsymbol{\mathrm{p}}) \right]  \\ \nonumber
    =&  -\mathbb{E}_{{\mathcal{X}}, {\mathcal{Y}},{\mathcal{P}}} \left[ \log P(\boldsymbol{\mathrm{y}}|\boldsymbol{\mathrm{x}}) \right]+\mathbb{E}_{{\mathcal{X}}, {\mathcal{Y}},{\mathcal{P}}} \left[ \log P(\boldsymbol{\mathrm{y}}|\boldsymbol{\mathrm{x}}, \boldsymbol{\mathrm{p}}) \right]
    \\ \nonumber
    =&  -\mathbb{E}_{{\mathcal{X}}, {\mathcal{Y}},{\mathcal{P}}}\left[ \log P(\boldsymbol{\mathrm{y}}|\boldsymbol{\mathrm{x}})-\log P(\boldsymbol{\mathrm{y}}|\boldsymbol{\mathrm{x}}, \boldsymbol{\mathrm{p}}) \right]\\ \nonumber
    =& -  \mathbb{E}_{{\mathcal{X}}, {\mathcal{Y}},{\mathcal{P}}}\left[\log\frac{p(\textbf{y}|\textbf{x})}{p(\textbf{y}|\textbf{x},\textbf{p})} \right]
\end{align}

We observe that $p(\textbf{y}|\textbf{x},\textbf{p})=\frac{p(\textbf{y},\textbf{p}|\textbf{x})}{p(\textbf{p}|\textbf{x})}$. Now, plugging it into Eq. \ref{eq:thm2-eq1}, we have
\begin{align}\label{eq:mutual}
    & -  \mathbb{E}_{{\mathcal{X}}, {\mathcal{Y}},{\mathcal{P}}}\left[\log\frac{p(\textbf{y}|\textbf{x})}{p(\textbf{y}|\textbf{x},\textbf{p})} \right]\\ \nonumber
    =& -  \mathbb{E}_{{\mathcal{X}}, {\mathcal{Y}},{\mathcal{P}}}\left[\log\frac{p(\textbf{p}|\textbf{x})p(\textbf{y}|\textbf{x})}{p(\textbf{y},\textbf{p}|\textbf{x})} \right].
\end{align}
Note that Eq.~\ref{eq:mutual} is also the definition of the \textbf{\textit{conditional mutual information}} $I(\textbf{Y};\textbf{P}|\textbf{X})$, which should always be \textbf{non-negative} \cite{cover1999elements}. Here, we will still provide further proof.

We can apply Jensen's Inequality, 
\begin{align}\label{eq:Jensen}
    &-  \mathbb{E}_{{\mathcal{X}}, {\mathcal{Y}},{\mathcal{P}}}\left[\log\frac{p(\textbf{p}|\textbf{x})p(\textbf{y}|\textbf{x})}{p(\textbf{y},\textbf{p}|\textbf{x})} \right] \\ \nonumber
     \overset{(a)}{\geq} & -  \log\mathbb{E}_{{\mathcal{X}}, {\mathcal{Y}},{\mathcal{P}}}\left[\frac{p(\textbf{p}|\textbf{x})p(\textbf{y}|\textbf{x})}{p(\textbf{y},\textbf{p}|\textbf{x})} \right] \\ \nonumber
    = &-  \log\mathbb{E}_{{\mathcal{X}}}\mathbb{E}_{{\mathcal{P}}, {\mathcal{Y}|{\mathcal{X}}}}\left[\frac{p(\textbf{p}|\textbf{x})p(\textbf{y}|\textbf{x})}{p(\textbf{y},\textbf{p}|\textbf{x})} \right] \\ \nonumber
    \overset{(b)}{=}& - \log 1
    = 0.
\end{align}
Here, Eq. \ref{eq:Jensen}(b) is because,
\begin{align}\label{eq:because}
    &\mathbb{E}_{{\mathcal{P}}, {\mathcal{Y}|{\mathcal{X}}}}\left[\frac{p(\textbf{p}|\textbf{x})p(\textbf{y}|\textbf{x})}{p(\textbf{y},\textbf{p}|\textbf{x})} \right] \\ \nonumber
    =& \sum_{ \textbf{y},\textbf{p}|\textbf{x}\in{\mathcal{P}}, {\mathcal{Y}|{\mathcal{X}}} } p(\textbf{y},\textbf{p}|\textbf{x}) \frac{p(\textbf{p}|\textbf{x})p(\textbf{y}|\textbf{x})}{p(\textbf{y},\textbf{p}|\textbf{x})} \\ \nonumber
    \overset{(a)}{=}& \sum_{ \textbf{y},\textbf{p}|\textbf{x}\in{\mathcal{P}}, {\mathcal{Y}|{\mathcal{X}}} } p(\textbf{p}|\textbf{x})p(\textbf{y}|\textbf{x}) \\ \nonumber
    =&  \sum_{ \textbf{p}|\textbf{x}\in{\mathcal{P}} {|{\mathcal{X}}} } p(\textbf{p}|\textbf{x})\sum_{ \textbf{y}|\textbf{x}\in{} {\mathcal{Y}|{\mathcal{X}}} }  p(\textbf{y}|\textbf{x}) \\ \nonumber
    =& 1.
\end{align}

The right term of (a) Eq. \ref{eq:because} can be interpreted as this: we denote a new random variable $\mathbf{Z}: p(\textbf{z})= p(\textbf{p}|\textbf{x})p(\textbf{y}|\textbf{x}), \textbf{z}\in {\mathcal{P}}, {\mathcal{Y}|{\mathcal{X}}}$. This is the outer product distribution which assigns probability $ p(\textbf{p}|\textbf{x})p(\textbf{y}|\textbf{x})$ to each $(\textbf{p},\textbf{y}|\textbf{x})$.
Hence, the right term of (a) in Eq. \ref{eq:because} becomes the sum of the probabilities of all possible outcomes of a random variable $\textbf{Z}$, which has to be 1.

Now, we go back to the inequality Eq. \ref{eq:Jensen}(a). Apparently, when $p(\textbf{y},\textbf{p}|\textbf{x})=p(\textbf{p}|\textbf{x})p(\textbf{y}|\textbf{x})$, the equality holds. Therefore, we can conclude that $H(\boldsymbol{\mathrm{Y}}|\boldsymbol{\mathrm{X}})- H(\boldsymbol{\mathrm{Y}}|\boldsymbol{\mathrm{X}}, \boldsymbol{\mathrm{P}})\geq 0$, and the equality holds \textit{if and only if} $\mathbf{P}$ and $\mathbf{Y}$ are conditionally independent given $\mathbf{X}$. The proof is completed.

\end{proof}

\section{More Discussion on Siamese Network}
\begin{figure*}[!t]
    \centering
    \includegraphics[width=0.8\linewidth]{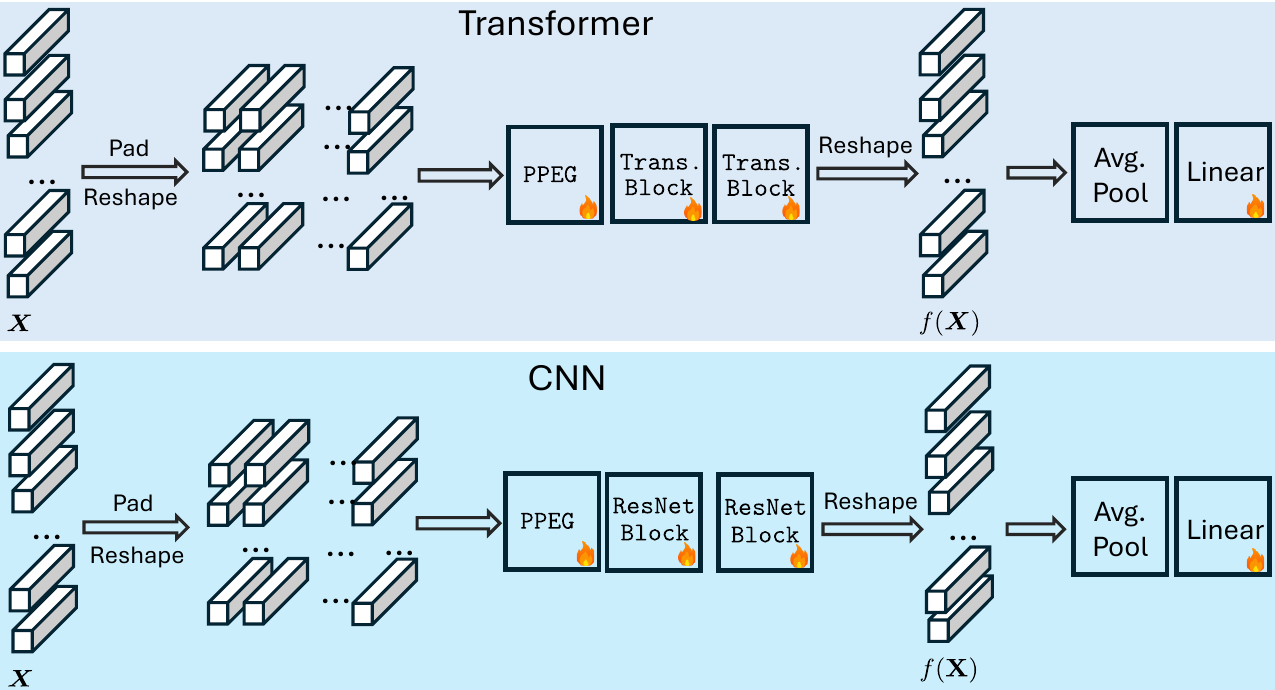}
    \caption{The network architecture of CNN ans Transformer used in the proposed method.}
    \label{fig:architect}
\end{figure*}

\subsection{Network Architecture}
The detailed network architectures of the two variants of the proposed method are shown in Fig.~\ref{fig:architect}. 

\noindent\textbf{Ours [Trans.]} Given extracted features from a bag $\mathbf{X}\in \mathbb{R}^{n\times d}$, we first apply a \texttt{MLP} for reducing its dimensions. Specifically, the MLP is able to be presented as: 
\begin{align}
\mathbf{X}\in \mathbb{R}^{n\times d}\xrightarrow{\texttt{g}(\cdot)}    \mathbf{X}\in \mathbb{R}^{n\times d/2}\xrightarrow{\texttt{g}(\cdot)} 
\mathbf{X}\in \mathbb{R}^{n\times 128},
\end{align}
where $\texttt{g}(\cdot)$ denotes a linear layer with a ReLU activation function. This operation can substantially reduce the computational cost.
Afterward, we perform \texttt{squaring} and apply \texttt{PPEG} as described in TransMIL \cite{shao2021transmil}. Then, we \texttt{flatten} the positioned features and feed them into two transformer blocks, and we obtain $f(\mathbf{X})$ or $f(S[\mathbf{X}])$ here. Further, we apply \texttt{Average Pooling} for aggregating all instance features for the final bag-level classification.  

\noindent\textbf{Ours [CNN]} Similar to \textbf{Ours [Trans.]}, we first apply \texttt{MLP}, \texttt{squaring}, and \texttt{PPEG} to obtain the positioned instance features. Then, we apply two residual blocks \cite{he2016deep}, where each blocks have the same input channel and output channel of 128. 
We obtain the block from \texttt{pytorch} official implementation (\url{https://github.com/pytorch/vision/blob/main/torchvision/models/resnet.py}).
Subsequently, we \texttt{flatten} the features and apply \texttt{average pooling} again for the final bag-level classification.

\subsection{1D CAM Derivation}
Class activation mapping (CAM)~\cite{zhou2016learning} is the most commonly used method to visualize the most discriminative areas that the model focuses on to make its decision. Typically, the CAM is computed for any network ended with a global average pooling and linear classifier, which is the primary reason why we use global average pooling for both CNN and Transformer. Below, we derive the 1-dimensional CAM in our case. First, we reshape the output from either our CNN or Transformer as a 2D matrix with $L$ dimensions: $F \in \mathbb{R}^{n \times L}$, where $n$ is the number of tiles in an WSI. We use subscript $i$ to index the feature vector corresponding to the $i$-th instance: $F_i \in \mathbb{R}^{l}$. The predicted classification logit for the $c$-th class $\texttt{logit}_c$ reads 
\begin{align*}
    \texttt{logit}_c &= \underbrace{\sum_{l=1}^L w_l^c \cdot\underbrace{\frac{1}{n}\sum_{i=1}^n F_i}_{\text{average pooling}}}_{\text{linear classifier}} \\
    &= \frac{1}{n}\sum_{i=1}^n \sum_{l=1}^L w_l^c F_i, 
\end{align*}
where $w_l^c$ is the $l$-th unit corresponding to the $c$-th class in the weight of the linear classifier. CAM is then defined as 
\begin{align*}
    \texttt{CAM}_i^c = \sum_{l=1}^L w_l^c F_i,
\end{align*}
which is the importance of the activation for the $i$-th instance leading to the classification of an WSI to class $c$.  

We provide the simplest solution to visualize the importance of each tile contributing to the final classification of an WSI via CAM. However, plain CAM relies on the global average pooling to obtain the importance. In a more general cases, where the network architecture may not contain global average pooling, the importance of each tile can also be obtained by using other variants of CAM, \textit{e.g.,} Grad-CAM~\cite{selvaraju2017grad}, Grad-CAM++~\cite{chattopadhay2018grad}, etc. However, this is not the main focus of this paper, hence we leave it for future work.

\subsection{Computational Complexity}
It is obvious that the Siamese network introduces additional computational complexity by having an additional branch compared to MIL methods. However, we argue that the overall computational burden in practice is still acceptable. In practice, most MIL methods do not effectively leverage parallel computation in modern deep learning training, as they typically set the batch size to 1 due to the varying number of instances. However, it is easy to fit a way larger number of bags/instances to most modern GPUs for better parallel acceleration. For the proposed Siamese network solution, we can easily stack the instances before and after shuffling to form a batch size of 2, as they have the same number of instances. By leveraging this parallel acceleration on GPUs, the computation of the proposed method can be as fast as the MIL methods with a single network branch. As shown in Table \ref{tab:time-cost}, we also empirically validate this assumption. We set the input bag with a shape $\mathbb{R}^{10000\times 768}$, which 
is aligned with the output feature dimension of the Swin-VIT extractor. The results illustrate that employing parallel acceleration can reduce the time cost by 41\% and 34\% for two of our proposed architectures, respectively. \textit{This makes the practical computation complexity of our dual-branch Siamese network is on par with previous MIL methods. }

\begin{table}[]
\centering
\caption{Time complexity (in \textit{ms}) per iteration by different strategies, benchmarked using input bags containing $10,000$ instances, each with $768$ dimensions: $\bm{X} \in \mathbb{R}^{10000\times 768}$. The single branch reduces to the same case of standard MIL schemes with only one branch of the neural network; whereas our Siamese solution requires a dual-branch network architecture. Dual-branch \textbf{w/o} parallelization denotes the training with a batch size of 1 as in standard MIL training; whereas  Dual-branch \textbf{w/o} parallelization denotes the training with a batch size of 2. }
\label{tab:time-cost}
\resizebox{0.9\linewidth}{!}{%
\begin{tabular}{ccc}\toprule
Strategy & \textbf{Ours [Trans.]} & \textbf{Ours [CNN]} \\ \midrule
{\color{gray} single-branch} & {\color{gray} 3.9} & {\color{gray} 1.5} \\ \midrule
\makecell{Dual-branch\\\textbf{w/o} Parallelization} & 7.8 & 2.9 \\ \midrule
\makecell{Dual-branch\\\textbf{w/} Parallelization} & 4.6 & 1.9 \\ \bottomrule
\end{tabular}%
}
\end{table}

\subsection{Additional Ablation Analysis on CTransPath Features}
Please refer to Table \ref{table:aba_ctrans} for the results, which show a 0.5\%-1.2\% gain obtained by the proposed model.

\begin{table}[!t]
\centering
\caption{Ablation Analysis on features extracted by CtransPath.}
\label{table:aba_ctrans}
\resizebox{0.99\linewidth}{!}{%
\begin{tabular}{cccccc|c}\toprule
Dataset               & Network                      & $\mathcal{L}_{\text{Equv}}$ & acc   & f1    & AUC   & Avg. 3 Metrics  \\ \midrule
\multirow{4}{*}{C16}  & \multirow{2}{*}{Trans.} & \ding{55}    & 95.89 & 95.62 & 97.42 & 96.31 \\
                      &                              & \ding{51}     & 96.64 & 96.39 & 98.00 & \textbf{97.01} \\ \cline{2-7}
                      & \multirow{2}{*}{CNN}         & \ding{55}     & 95.21 & 94.82 & 96.67 & 95.57 \\
                      &                              & \ding{51}   & 96.25 & 95.99 & 98.10 & \textbf{96.78} \\ \midrule
\multirow{3}{*}{TCGA} & \multirow{2}{*}{Trans.} & \ding{55}     & 94.53 & 94.57 & 97.66 & 95.59 \\
                      &                              & \ding{51}    & 95.20 & 95.19 & 97.99 & \textbf{96.13} \\ \cline{2-7}
                      & \multirow{2}{*}{CNN}         & \ding{55}    & 94.43 & 94.43 & 96.99 & 95.28 \\
\multicolumn{1}{l}{}  &                              & \ding{51}    & 95.11 & 95.10 & 97.55 & \textbf{95.92} \\ \bottomrule
\end{tabular}%
}

\vspace{-0.1cm}
\end{table}

\section{A Complete Justification from OT Theory}
To make this justification complete and self-contained, we start with the introduction of the forward OT formulation of cracking instance jigsaw puzzles in Sec.~\ref{sec:forward_ot} and then introduce our inverse OT formulation and solution in Sec.~\ref{sec:inverse_ot}. 

\subsection{Forward OT Formulation}\label{sec:forward_ot}
Our instance jigsaw puzzles can be formulated as a (forward) OT problem by minimizing the Wasserstein distance (\textit{e.g.,} Earth Mover's Distance) between coordinates associated with the shuffled instances ($\bm{\mathrm{p}}'$) and those associated with the unshuffled bag ($\bm{\mathrm{p}}$):
\begin{align*}\label{eq:emd}
    \text{EMD}(\boldsymbol{\mathrm{P}}, \boldsymbol{\mathrm{P}}') &= \min\limits_{\bm{\mathrm{T}} \geq 0} \sum_{i, j} \bm{\mathrm{T}}_{ij} \bm{\mathrm{C}}^{ij}\\
    \text{subject to}  \ \sum_i \bm{\mathrm{T}}_{ij} &= \frac{1}{n}, \ \sum_j \bm{\mathrm{T}}_{ij} = \frac{1}{n}  
\end{align*}
where $\bm{\mathrm{T}}$ is the transport flow matrix, and $\bm{\mathrm{C}}$ is the known cost matrix (\textit{e.g.,} a quadratic cost: $\|p_i - p'_j\|_2^2$). 

This forward OT formulation aims at solving the unkown optimal transport plan (\textit{i.e.,} $\bm{\mathrm{T}}_\#$) that can restore the instance ordering from its random arrangement. However, solving this forward OT problem is non-trivial, as it necessitates iterative updates (\textit{e.g.,} Sinkhorn updates~\cite{cuturi2013sinkhorn,peyre2019computational}). This makes solving the forward OT problem computationally challenging, particularly when deep neural networks are involved in our case. 

\subsection{Inverse OT Formulation (Proof of Theorem 3)}\label{sec:inverse_ot}
Our key insight is that solving instance jigsaw puzzles may not require an optimal plan; rather, non-optimal plans can also achieve the same objective. Therefore, we can consider the simplest plan $\Tilde{\bm{\mathrm{T}}}$, \textit{i.e.,} the inverse shuffling operation $\mathcal{S}^{-1}$ (or equivalently, $\mathbf{P}_\sigma^\top$ in a matrix form). 

\begin{proof}
The inverse optimal transport (OT) problem can be formulated as solving the following optimization problem:
\begin{align*}
    &\min\limits_{\theta} \mathcal{L}(\Tilde{\bm{\mathrm{T}}}, \bm{\mathrm{T}}_\#[c_\theta]) \\
    \text{subject to} & \quad  \bm{\mathrm{T}}_\#[c_\theta] = \min\limits_{\bm{\mathrm{T}} \geq 0} \sum_{i, j} \bm{\mathrm{T}}_{ij} \mathrm{c}_{\theta}^{ij},
\end{align*}
where $c_\theta$ is the parameterized cost function, and $\mathcal{L}(\cdot)$ is a loss function, \textit{e.g.,} MSE. $\Tilde{\bm{\mathrm{T}}}$ and $\bm{\mathrm{T}}_\#$ are the observed and optimal transport plan.
Some common choices of the loss function $\mathcal{L}$ are mean-squared loss (MSE) and Kullback–Leibler (KL) divergence. However, solving the inverse OT problem requires to solve the forward OT problem to obtain the optimal transport plan, which is intractable in our case. Hence, we approximate the optimal transport plan $\mathbf{T}_\#$ with the observed plan $\Tilde{\mathbf{T}}$. The above inverse OT objective is simplified to: 
\begin{align*}
    &\min\limits_{\theta} \sum_{ij} \Tilde{\mathbf{T}}_{ij}c_\theta^{ij},
\end{align*}
where $c_\theta^{ij} = c_\theta(\mathbf{x}_i, \mathbf{x}'_j)$. In the most naive case, $c_\theta(\mathbf{x}_i, \mathbf{x}'_j)$ should be a neural network that takes as input both $\mathbf{x}_i$ and $\mathbf{x}'_j$. Instead, we consider penalizing the $\mathrm{L}_2$ norm between $\mathbf{x}_i$ and $\mathbf{x}'_j$ in the embedding space: $\|f_\theta(\mathbf{x}_i) - f_\theta(\mathbf{x}'_j) \|_2^2$, where $f_\theta$ is a neural network.

If we further replace the observed plan $\Tilde{\mathbf{T}}_{ij}$ with the inverse shuffling operation in a matrix form $(\mathrm{P}_{\sigma}^\top)_{ij}$ (see Sec.~\ref{sec:ap:1}), the above minimization reduces to 
\begin{align*}
    \min\limits_{\theta} \sum_{ij} (\mathrm{P}_{\sigma}^\top)_{ij} \|f_\theta(\mathbf{x}_i) - f_\theta(\mathbf{x}'_j) \|_2^2.
\end{align*}
Note that $\mathrm{P}_{\sigma}^\top$ is an orthonormal matrix with exactly one elements equal to 1 at each row and column. Hence, $\sum_{i} (\mathrm{P}_{\sigma}^\top)_{ij} = \sum_{j} (\mathrm{P}_{\sigma}^\top)_{ij} = 1$ instead of $1 / 
 n$. However, this does not affect the results of minimization. Putting everything in a matrix form, we have 
\begin{align*}
    \min\limits_{\theta} \frac{1}{2n} \|f_\theta(\mathbf{X}) -   \mathbf{P}_\sigma^\top f_\theta(\mathbf{X}_\sigma) \|_2^2,
\end{align*}
where $\mathbf{X}_\sigma = \mathbf{P}_\sigma \mathbf{X}$. Equivalently, we have 
\begin{align*}
    \min\limits_{\theta} \frac{1}{2n} \|\mathbf{P}_\sigma f_\theta(\mathbf{X}) -  f_\theta(\mathbf{X}_\sigma) \|_2^2.
\end{align*}
Replacing the permutation matrix $\mathbf{P}_\sigma$ with the corresponding operator $\mathcal{S}^{-1}$, the above equation reduces to the same as our shuffling equivalence regularization loss. This concludes the proof.
\end{proof}

\section{Additional Implementation Details}

\paragraph{Reproducibility.} The code will be made publicly available upon acceptance. \textbf{The core code is provided in the Supplementary file.}

\subsection{Implementation Hyperparameters}
We adopt the default training setup for baseline models. For our proposed models, we use the AdamW optimizer \cite{loshchilov2017decoupled} with an initial learning rate of 5e-4 and a weight decay of 1e-4 for all experiments. Each model is trained for 200 and 20 epochs with a batch size of 1 for WSI classification and survival prediction, respectively. All experiments are implemented in PyTorch (v1.13) \cite{paszke2019pytorch} and performed on an NVIDIA A100 GPU with 40 GB memory.
We present the hyperparameters of all baselines as well as our model in Table \ref{parameter}.
\begin{table*}[htbp]
    \centering
    \caption{Hyperparamters used in the experiments. Here, \texttt{Cosine$^*$} denotes cosine decay with 20 epoch linear warmup from 1e-5. \texttt{AMP} denotes automatic mixed precision. }
    \resizebox{0.99\textwidth}{!}{
    \begin{tabular}{c|l|l|l|l|l|l|l|l|l}\toprule
        \multicolumn{1}{c|}{\textbf{Setting}} & \textbf{AB-MIL} &  \textbf{DS-MIL}& \textbf{DTFD-MIL} & \textbf{Trans-MIL}  & \textbf{ILRA-MIL}  & \textbf{DGR-MIL }  & \textbf{MHIM-MIL } & \textbf{AC-MIL} & \textbf{Ours}\\ \hline
        Optimizer & Adam    &    Adam  &      Adam  &  Radam  & Adam & SGD & Adam & Adam &AdamW  \\
        Learning rate  &  1e-3     &    1e-4  &   1e-4  &   2e-4  & 1e-4 & 5e-4 & 2e-4 &1e-4/2e-4 &5e-4  \\
        Weight decay  &    0.005   &   5e-3 &   1e-4 & 1e-5 & 1e-4 & 1e-4& 1e-5 &1e-4 &1e-4\\               
        Scheduler  &   Cosine*& Cosine &  MultiStepLR &   LookAhead~\cite{lookahead}  & Cosine & Cosine*& Cosine& Cosine &  LookAhead~\cite{lookahead}\\ 
        Loss  &    $ \mathcal{L}_{bce}$ &    $ \mathcal{L}_{bce}$   &    $ \mathcal{L}_{bce}$ +  Tier-2 loss &  BCE   &  $ \mathcal{L}_{bce}$   & $ \mathcal{L}_{bce}, \mathcal{L}_{tri},\mathcal{L}_{div}$&$ \mathcal{L}_{bce},\mathcal{L}_{con}$  & $ \mathcal{L}_{bce},\mathcal{L}_{p}, \mathcal{L}_{d}$ & $ \mathcal{L}_{bce}, \mathcal{L}_{Equv} $ \\ 
        other  &  None    &   Droppath = 0.2 &   grad. clip = 5 & AMP & Xavier initialize & Warmup & None & None &  None   \\ \bottomrule
    \end{tabular}
    }
    \label{parameter}
    
\end{table*}

\subsection{Dataset for WSI classification}
\noindent\textbf{CAMELYON16 dataset} is a publicly available collection of WSIs designed to detect metastatic breast cancer in lymph node tissue. It comprises 399 WSIs of lymph node specimens, officially divided into a training set of 270 samples and a test set of 129 samples. Each WSI is associated with a binary label, annotated by expert pathologists, indicating the presence or absence of metastatic cancer. In addition, detailed region-level annotations are provided for cancerous tissue within the WSIs. We employ a threshold-based preprocessing method to filter background information for patch extraction~\cite{zhang2022dtfd,otsu1975threshold}. Each WSI image is cropped into non-overlapping patches of size 256 × 256, resulting in approximately 4.61 million patches at ×20 magnification, with an average of 11,555 patches per slide. 

\noindent\textbf{TCGA-NSCLC dataset} is also a public dataset for classifying two subtypes of lung cancer (lung squamous cell carcinoma and adenocarcinoma). It includes a total of 1,037 WSIs. Following the preprocessing protocol outlined in CAMELYON16~\cite{zhang2022dtfd,otsu1975threshold}, approximately 13.83 million patches were extracted at ×20 magnification. On average, each WSI yielded 13,335 patches per sample.

We also want to highlight that with three different feature extractors, our experiments are three times the scale compared to previous studies~\cite{zhang2022dtfd,shao2021transmil}.

\subsection{Dataset for WSI Survival Analysis}

Both of the following datasets employ a similar pre-processing in that all patches were extracted at ×20 magnification.

\noindent\textbf{TCGA-LUAD dataset} is a subset of TCGA-NSCLC dataset that only has a subtype of lung cancer (lung adenocarcinoma). Following the dataset (i.e. \texttt{csv} file from  \url{https://github.com/mahmoodlab/Panther}) provided by \cite{song2024morphological}, we processed a subset of the dataset, comprising 463 WSIs from 412 patients.

\noindent\textbf{TCGA-BRCA Dataset} is a comprehensive collection curated for the study of Breast Invasive Carcinoma. Following the dataset provided in \cite{jaume2024modeling}, we processed this dataset to include 931 WSIs from 871 patients.

\subsection{Additional Implementation of WSI Survival Analysis}
Our implementation strictly adheres to \cite{jaume2024modeling}, and we use disease-specific survival (DSS). The training framework (e.g., Loss function, the data partition) is implemented based on their repositories: \url{https://github.com/mahmoodlab/SurvPath}. The features we used is extracted by \textbf{UNI}~\cite{chen2024uni} extractor.

\section{Statistical Test}
\begin{figure}[!t]
    \centering
    \includegraphics[width=0.9\linewidth]{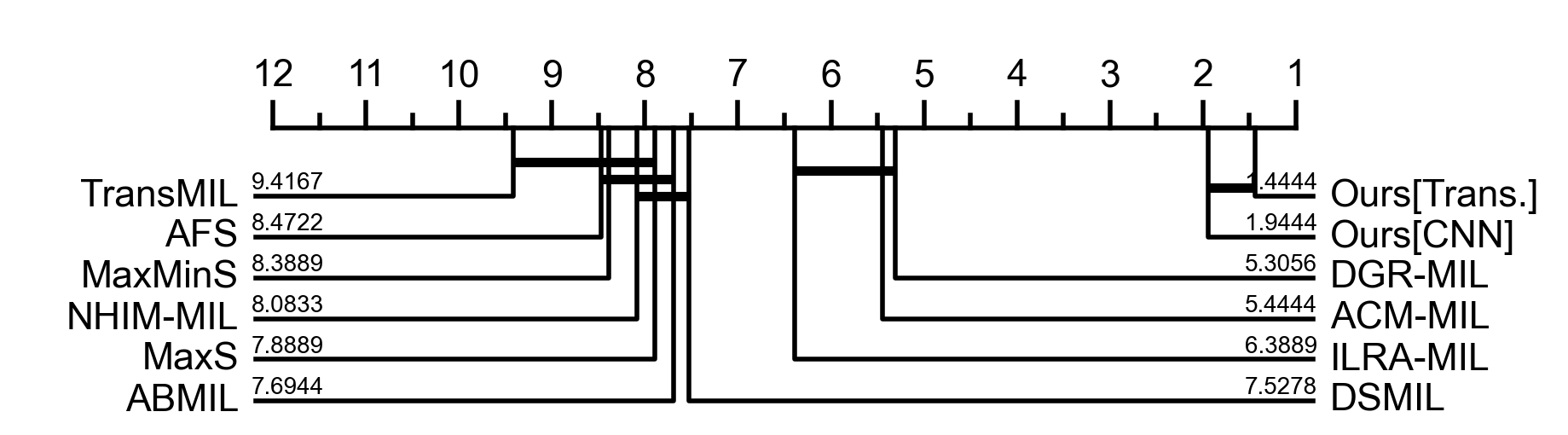}
    \caption{Critical difference diagram based on the Wilcoxon signed-rank test, where the number indicates the average ranks ($\uparrow$: the higher the better). Methods connected by a single thick line show no statistically significant differences.}
    \label{fig:Critical}
\end{figure}

Following the recommendation by \cite{demvsar2006statistical}, we utilize the Wilcoxon signed-rank test to compare two classifiers on a single dataset. For summarizing the performance across multiple datasets and feature extractors, we employ a critical difference diagram. In Fig.~\ref{fig:Critical}, we present the critical difference diagram with a significance level of $\alpha = 0.05$ (which is a common threshold in hypothesis testing; if the p-value is lower than this value, we treat the performance of two classifiers are significantly different), illustrating that two of our proposed models statistically outperform previous MIL baselines.

\end{document}